\documentclass[usegraphicx,usenatbib,useAMS]{mn2e}

\usepackage{times} 
\usepackage{aas_macros}     
\usepackage{amssymb}        

\newcommand{\gsim}{\gtrsim} 

\newcommand{\hMsol}{h^{-1}M_\odot}

\title[Lyman-break galaxies in $\Lambda$CDM]{The nature and descendants of Lyman-break galaxies
in the $\Lambda$CDM cosmology.}

\author[J. E. Gonz\'{a}lez et al.]{Juan
E. Gonz\'{a}lez$^{1,2}$\thanks{E-mail: jegonzal@eso.org},
C. G. Lacey$^{1}$, C. M. Baugh$^{1}$, C. S. Frenk$^{1}$,
A. J. Benson$^{3}$ \\
$^{1}$Institute for Computational Cosmology,
Department of Physics, Durham University, South Road, Durham DH1 3LE,
UK.\\ 
$^2$European Southern Observatory, Karl-Schwarzschild-Str. 2, 
85748 Garching, Germany.\\
$^{3}$Mail Code 350-17, California Institute of Technology, Pasadena, CA 91125, USA}

\begin{document}

\date{}

\pagerange{\pageref{firstpage}--\pageref{lastpage}} \pubyear{2009}

\maketitle
 
\label{firstpage}

\begin{abstract}
We predict the formation histories, properties and descendants of
Lyman-break galaxies (LBGs) in the $\Lambda$CDM cosmology. In our
model, which incorporates a top-heavy IMF in starbursts, we find that
most LBGs are starbursts triggered by minor mergers of galaxies. The
duration of the LBG phase is predicted to be quite short, $\sim
20-60$~Myr. We investigate the distributions of stellar and halo
masses and morphologies for bright ($L_{UV} > L^{*}_{UV}$) and faint
($L_{UV} > 0.1 L^{*}_{UV}$) LBGs at $z=3$, $z=6$ and $z=10$ (where we
classify LBGs according to their rest-frame UV luminosities relative
the observed characteristic luminosity $L^{*}_{UV}$ at $z \approx
3$). Bright LBGs at $z=3$ are predicted to have median stellar masses
$\sim 1 \times 10^9 \hMsol$ and host halo masses $\sim 3 \times
10^{11} \hMsol$, and to be typically mildly disk-dominated in stellar
mass. On the other hand, faint LBGs at $z=10$ are predicted to have
median stellar masses of only $\sim 1 \times 10^7 \hMsol$ and host
halo masses $\sim 2 \times 10^{10} \hMsol$, and to be generally
bulge-dominated. Bright LBGs at $z=3$ evolve to present-day galaxies
with median stellar mass $\sim 5\times10^{10} h^{-1}M_{\odot}$
(comparable to the Milky Way), consisting of roughly equal numbers of
disk- and bulge-dominated systems, and hosted by halos with median
mass $\sim 2\times10^{13} h^{-1}M_{\odot}$ (corresponding to
medium-size galaxy groups). 
The model predicts that
40\% of Milky Way mass galaxies at the present-day have a bright LBG
progenitor in the redshift range $3<z<4$, while 95\% have a faint LBG
progenitor in the same redshift range, and 7\% have a faint LBG
progenitor at $10<z<11$. With our multiwavelength model, we also
investigate the overlap between the LBG population and that of
submillimetre selected galaxies (SMGs); at $z=3$, only $\sim 1\%$ of
bright LBGs are also predicted to also be bright SMGs (with an $850
\mu$m flux in excess of 5~mJy).
\end{abstract}


\section{Introduction}
%

High redshift galaxy populations identified using the Lyman-break
drop-out technique have played a pivotal role in the development of
our understanding of the global star formation history of the Universe
\citep{Madau96,Ste99}. The star formation rate density inferred from
measurements of the galaxy luminosity function in the rest frame
ultraviolet (UV) displays a broad peak around redshift $\sim 2$, prior
to a steep decline towards the present day \citep{Lilly96}.  This type
of behaviour, which corresponds to a gradual build up of the global
stellar mass in galaxies, arises naturally in hierarchical structure
formation models, as shown by \citet{Cole94} and confirmed in a range
of calculations with varying galaxy formation physics
\citep[e.g.][]{Baugh98,Somerville01,Springel03}. In general terms, the
star formation rate density is set by the abundance of haloes as a
function of mass which are in place at a given epoch, the ability of
gas to cool within these haloes and the effectiveness of supernovae at
reheating gas which has cooled. The star formation density peaks
around $z \sim 2$ because gas can cool efficiently at high densities
and stays in the cold phase because there is, for the first time, a
large population of haloes in which supernova heating is
ineffective. In this paper we investigate the properties of
Lyman-break galaxies (hereafter LBGs) in a hierarchical galaxy
formation model and show how they are related to galaxies in the
present day Universe.

The Lyman break technique is based on observing a break bluewards of
the Lyman limit at 912\AA\ in the rest-frame spectral energy
distribution of a galaxy. The flux at these wavelengths is expected to
be small for a number of reasons; i) few photons can be produced by
stars at these short wavelengths, ii) such energetic photons can be
absorbed by the atomic hydrogen in the interstellar medium within the
galaxy, and iii) the photons can also be absorbed by intergalactic
neutral hydrogen along the line of sight from the galaxy to the
Earth. Galaxies identified with this technique are known as
Lyman-break galaxies (hereafter LBGs,\citealt{Ste92,Ste96}).  The
Lyman-break technique uses multi-band photometry. By using a suitable
set of filters, galaxies in a particular range of redshifts can be
isolated.  Ground-based observations are restricted to applying the
Lyman-break technique for galaxies at $z>2.5$. For this redshift
range, the absorption edge, which occurs in the far-UV, is shifted to
wavelengths that are not blocked by the Earth's atmosphere. The
multi-band photometry is also used to make additional colour
selections to minimize the contamination from foreground objects (both
stars and lower redshift galaxies).  LBGs are selected to be red in
the shorter wavelength colour, due to the presence of the break, and
blue in the longer wavelength colour, to show they are star-forming
galaxies. Hot, high mass stars with short lifetimes tend to dominate
the emission in the UV. Such stars spend a significant fraction of
their lifetime within dense giant molecular clouds and their light is
heavily extincted. Hence in order to infer a galaxy's star formation
rate from its UV luminosity requires an uncertain and quite often
large correction for dust extinction.  Furthermore, an assumption
about the form of the initial mass function with which stars are
produced is also required in order to infer the star formation rate.

In this paper we use a semi-analytical approach to model the galaxy
formation process (see \citealt{Baugh06} and \citealt{Benson10} for
reviews). This type of model can be used to make predictions for the
properties of high redshift galaxies and can connect these objects to
their present day descendants.  This approach was first applied to the
Lyman-break population by \citet{Baugh98}, who showed that the model
of \citet{Cole94} could reproduce the observed surface density of
Lyman-break galaxies \citep[e.g.][]{Ste96}. \cite{Somerville01}
applied similar techniques to argue that bursts triggered by minor and
major galaxy mergers played a role in producing the LBG population.
In this paper we use the version of the Durham {\tt GALFORM} galaxy
formation simulation code published by \cite{Baugh05} to study LBGs
and their descendants. This model invokes a different form for the
stellar initial mass function in merger-driven starbursts, in order to
reproduce the observed counts and redshift distribution of galaxies
measured at sub-millimetre wavelengths.  \cite{Baugh05} showed that
this model also matches the observed LBG luminosity function at $z=3$,
as well as giving a reasonable match to the present day galaxy
population. \cite{Gonzalez09} extended the local tests of this model
to include the size-luminosity relation for disk-dominated
galaxies. The semi-analytical model naturally relates galaxies at high
redshift to their local counterparts.  In \cite{Gonzalez09b} we
followed the evolution of submillimetre galaxies (SMGs) and their
descendants. In this paper, we carry out a similar analysis in the UV
by studying the evolution and descendants of Lyman-break galaxies.  In
a companion paper to this one, \citet{Lacey10b} present model
predictions for the evolution of the far-UV luminosity functions over
the redshift range $z=3-20$, and make a detailed comparison of these
with current observational data. They also make predictions for the
amount of dust extinction, and for the evolution of a wide variety of
other properties of LBGs over a range of redshifts. The present paper
instead looks in more detail at the formation histories and lifetimes
of LBGs and the distributions of stellar and halo masses and
morphologies for LBGs at selected redshifts and luminosities, and
investigates the corresponding properties for their present-day
descendants.

This paper is laid out as follows. In Section 2, we outline the {\tt
GALFORM} galaxy formation model. In Section 3, we review the model
predictions for the luminosity function of LBGs, show examples of
galaxy merger histories which include LBGs, and investigate the
triggering and duration of the LBG phase. We also present the
probability that a present day galaxy or halo had an LBG progenitor.
In Section 4, we study different properties of the LBGs, such as
stellar and host halo masses, and morphologies at redshifts $z=3$,
$z=6$ and $z=10$, and discuss the dominant mode of star formation
behind the LBGs.  In Section 5, we examine the overlap between the LBG
and SMG populations.  Finally, we present a summary in Section 6.  We
quote magnitudes on the AB system and define the Hubble parameter as
$h=H_{0}/100$ km s$^{-1}$ Mpc$^{-1}$.

\section[]{The galaxy formation model}

We first give a brief overview of the Durham semi-analytical galaxy
formation code, {\tt GALFORM}, before discussing two aspects which are
of particular relevance for this paper, galaxy mergers and the form of
the IMF adopted in different modes of star formation. Full details of
{\tt GALFORM} can be found in \citet{Cole00} and \citet{Benson03}.
The model used here is described in full in \citet{Baugh05} and
\citet{Lacey08}.

\subsection{Basic Components}

Galaxies are assumed to form inside dark matter haloes, with their
subsequent evolution being controlled by the merger and accretion
history of the halo \citep{White78}. The physical processes modelled
in {\tt GALFORM} include: i) the hierarchical assembly of dark matter
haloes; ii) the shock heating and virialization of gas inside the
gravitational potential wells of dark matter haloes; iii) the
radiative cooling of the gas to form a galactic disk; iv) star
formation in the cool gas; v) the heating and expulsion of cold gas
through feedback processes such as stellar winds and supernovae; vi)
chemical evolution of gas and stars; vii) galaxy mergers within a
common dark halo as the result of the decay of a galaxy's orbit due to
dynamical friction; viii) the evolution of the stellar populations
using population synthesis models; ix) the extinction and reprocessing
of starlight by dust.

We use the same galaxy formation parameters adopted by \citet{Baugh05}
and \citet{Lacey08}. The model is set in a cold dark matter plus
cosmological constant cosmology, with present-day matter density
parameter $\Omega_{0}=0.3$, a cosmological constant
$\Omega_{\lambda}=0.7$, a baryon density $\Omega_{b}=0.04$ and a power
spectrum normalization given by $\sigma_{8}=0.93$.

In this paper one of our primary goals is to connect high redshift
galaxies with their present-day descendants. This requires a
prescription for building halo merger histories which is accurate over
a long interval in time.  To achieve this we use the Monte-Carlo
algorithm of \cite{Parkinson08}, which is a modified
version of that proposed by \cite{Cole00} and used in \cite{Baugh05}.
The \citeauthor{Parkinson08} recipe has been tuned to match the merger
histories extracted from the Millennium Simulation
(\citealt{Springel05}; see \citealt{Harker06} for a description of the
construction of merger trees from the simulation).  We note that
\citeauthor{Baugh05} did not make any predictions which depended on
the accuracy of the trees over long time intervals. Rather, they laid
down grids of halos at a range of redshifts to compute the luminosity
function of SMGs and other galaxies at these redshifts, and then, by
integration, the number counts of SMGs and the cosmic star formation
history. We have checked that using the \citeauthor{Parkinson08} tree
algorithm does not alter the predictions presented in
\citeauthor{Baugh05}

Dust extinction is expected to have a large effect on the far-UV
luminosities of LBGs. We calculate dust extinction in the model using
the approach described in \citet{Lacey10b}, and we refer the reader to
that paper for a discussion of the effects of dust extinction on
LBGs. In summary, the dust extinction is calculated using a
two-component dust model, with molecular clouds within which stars
form embedded in a diffuse dust medium mixed with the stars and
distributed in an exponential disk.  All of the predictions which we
show in this paper include the effects of dust extinction.

\subsection{Galaxy mergers and merger-driven starbursts}

When dark matter haloes merge, the galaxy in the most massive halo is
assumed to become the central galaxy in the new halo while the other
galaxies present become its satellites. In the standard form of {\tt
GALFORM}, we assume that central and satellite galaxies can form stars
from their cold gas reservoirs, but that only central galaxies can
continue to accrete cold gas.  (\citealt{Font08} explore a variant of
this model in which satellites are allowed to retain some fraction of
their hot halo, depending on their orbit within the main halo, with
gas allowed to cool onto the satellite disks.)

In the standard version of the model, mergers are assumed to happen
only between satellites and central galaxies within the same halo
(although \citealt{Kim09} considered a variant of this).
Bursts of star formation are assumed to be triggered by galaxy
mergers.  Two types of mergers are defined, major mergers and minor
mergers, according to whether the ratio of  mass of the smaller to
the larger galaxy exceeds the value of the parameter $f_{\rm ellip}$,
i.e. $M_2/M_1 > f_{\rm ellip}$.  Bursts are assumed to be triggered in
all major mergers, and also in minor mergers which satisfy $M_2/M_1 >
f_{\rm burst}$ and where the ratio of cold gas mass to stellar plus
cold gas mass in the larger galaxy exceeds $f_{\rm gas,crit}$.  The
parameter values adopted in the Baugh et.~al model are $f_{\rm
ellip}=0.3$, $f_{\rm burst}$=0.05 and $f_{\rm gas,crit}=0.75$.  In a
major merger, the stellar disks of the merging galaxies are assumed to
be transformed into a new stellar spheroid (resulting in a pure bulge
galaxy) and any gas present in the disks is converted into spheroid
stars in a burst. In minor mergers, the stellar disk of the larger
galaxy is preserved and the stars of the smaller, accreted galaxy are
added to its bulge (resulting in a disk plus bulge galaxy). If the
condition for a burst in a minor merger is satisfied, then the gas of
the smaller galaxy is converted into stars. If the condition is not
satisfied, any accreted gas is added to the gas disk of the central
galaxy. From this description of the possible scenarios following a
galaxy merger, it is readily apparent that mergers change the
morphology of galaxies in the model, driving the morphology towards
early types (bulge dominated), with the cooling of gas moving the
galaxy towards late types (disk dominated).

\subsection{The stellar IMF in quiescent and burst modes of star formation}

An important feature of the Baugh et~al. model is that stars are
assumed to form with different stellar initial mass functions (IMFs)
in different modes of star formation. Stars either form quiescently in
galactic disks or in bursts triggered by some galaxy mergers, as
explained in the subsection above.  In the case of quiescent star
formation, a standard solar neighbourhood IMF is adopted, following
the form proposed by \citet{Kennicutt83}, where the number of stars
produced depends on mass as ${\rm d}N/{\rm d}\ln m \propto m^{-x}$,
with $x=0.4$ for $m < 1 M_{\odot}$ and $x=1.5$ for $m > 1
M_{\odot}$. In the case of merger-driven starbursts, a top-heavy IMF
is assumed where $x=0$.  The use of a top-heavy rather than solar
neighbourhood IMF has three consequences; i) it increases the amount
of radiation emitted in the UV, as a larger proportion of massive
stars is produced, ii) there is higher yield of metals ($p=0.15$
compared with $p=0.023$ for the Kennicutt IMF), leading to more dust
and iii) the fraction of mass turned into stars which is recycled to
the interstellar medium is larger ($R=0.91$ compared with $R=0.41$ for
the Kennicutt IMF).  For a discussion of observational support for a
top-heavy IMF see \citet{Lacey08,Lacey10a}.

Finally, in closing this section we remind the reader that an
alternative galaxy formation scenario was presented by
\citet{Bower06}, who invoke a different mechanism to suppress the
formation of galaxies in massive haloes than that used by
\citeauthor{Baugh05} \citeauthor{Bower06} also invoke a standard IMF
in both modes of star formation.  Whilst this model gives a good match
to the observed evolution of the K-band luminosity function and the
inferred stellar mass function, it greatly underpredicts the number of
submillimetre galaxies and gives a  poorer match to the LBG
luminosity function than the Baugh et~al. model
\citep[see][]{Lacey10b}.  For further discussion of the differences
between these two models, see \citet{Almeida07}, \citet{Violeta09} 
and \citet{Gonzalez09}.

\section{LBG formation histories and the probability of having a LBG progenitor}

\begin{figure}
\includegraphics[width=8.4cm]{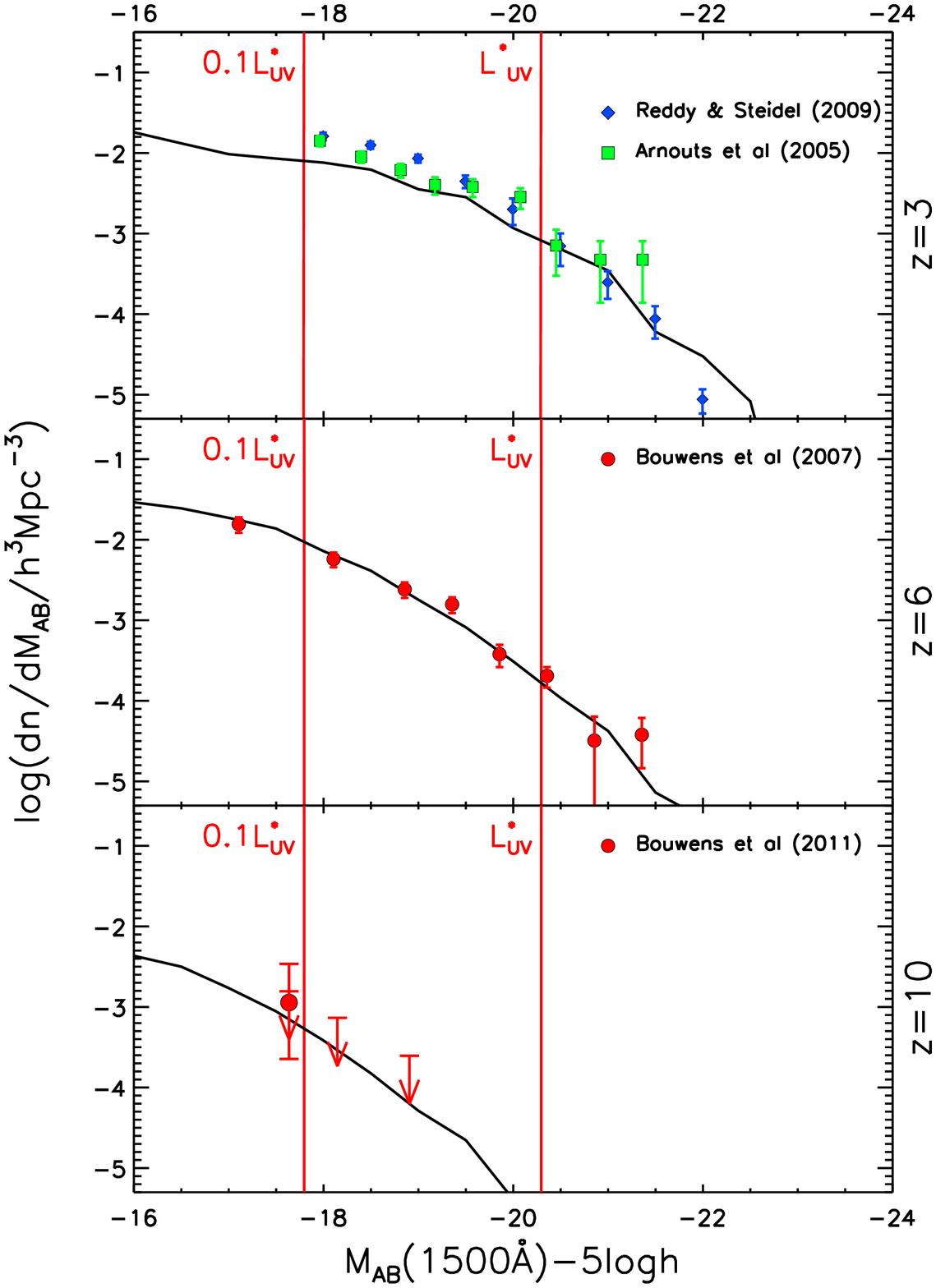}
\caption{ The galaxy luminosity function in the rest-frame UV
  (1500\AA) at $z=3$ (top panel), $z=6$ (middle panel) and $z=10$
  (bottom panel). The symbols with error bars show observational
  estimates taken from \citet{Arnouts05} and \citet{Reddy09} at $z=3$,
  \citet{Bou07a} at $z=6$ and \citet{Bou10} at $z=10$ as indicated by
  the legends. 
  The predicted luminosity functions include the effect of
  dust extinction and the observational estimates have not been
  corrected to remove extinction.  The vertical red lines indicate
  $L^{*}_{UV}$ and $0.1 L^{*}_{UV}$ (defined at z=3) which we use to
  construct samples of bright and faint LBGs respectively.}
\label{LF}
\end{figure}

\begin{figure*}
\includegraphics[width=15.5cm, bb=0 0 550 750]{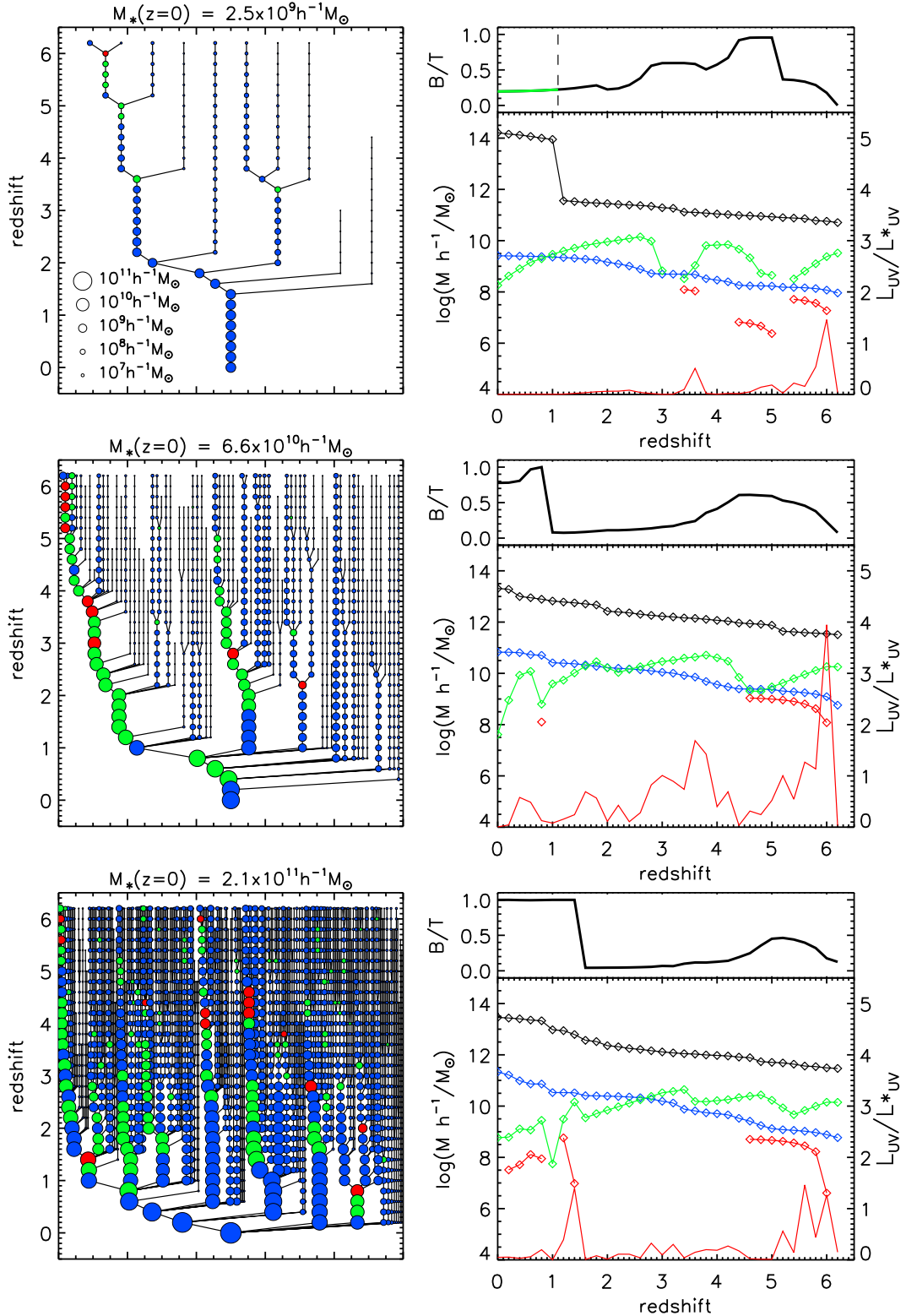}
\caption{ Left panels: galaxy merger trees for different $z=0$
galaxies.  From top to bottom, the stellar masses of the galaxies at
$z=0$ are $M_{*}=2.5\times10^{9}, 6.6\times10^{10}$ and $2.1
\times10^{11} h^{-1}M_{\odot}$ respectively. The size of the symbol is
proportional to the stellar mass as indicated by the key in the top
left panel. Green circles are drawn for galaxies with $L_{UV} > 0.1
L^{*}_{UV}$, red circles for $L_{UV} > L^{*}_{UV}$ and blue if $L_{UV}
< 0.1 L^{*}_{UV}$. Right panels: evolution with redshift of the galaxy
on the main progenitor branch. Top subpanel: evolution of the bulge to
total stellar mass ratio ($B/T$). The dashed vertical line indicates
the redshift at which the central galaxy (black line) becomes a
satellite galaxy (plotted using a green line) after a halo
merger. Main right subpanel: the symbols show the evolution of the
stellar mass (blue), cold gas mass (green), host dark matter halo mass
(black) and the stellar mass formed in ongoing bursts (red) in units
according to the left-hand label.  The red line shows the evolution of
the ratio $L_{UV} / L^{*}_{UV}$ with redshift, in units according to
the right-hand axis. Note that $L^{*}_{UV}$ is the observationally
determined reference value at $z=3$ (\citealt{Ste99}).  }
\label{GMT}
\end{figure*}

In this section we present some basic predictions for LBGs in the
Baugh et~al.  model. We first compare the model predictions for the
luminosity function in the UV at $z=3$, $z=6$ and $z=10$ with
observations, and use this to set up definitions of bright and faint
LBGs. (For a more detailed comparison with observed LBG luminosity
functions see \citealt{Lacey10b}.)  We then present some illustrative
galaxy formation histories for present day galaxies which had LBG
progenitors. Finally we show the probability that a present-day galaxy
had an LBG progenitor and that a present-day halo hosted an LBG.

\subsection{The UV luminosity function: defining bright and faint Lyman-break galaxy samples.}

As outlined in the Introduction, the Lyman-break technique relies on
colour-colour selection to identify galaxies within a particular
redshift range, which is set by the redshifted Lyman-break spectral
feature falling between two of the filters used to image the
galaxies. In the model we can make predictions for the galaxy
population at any desired redshift by construction and so there is no
need to apply any colour selection. Hence, we will use the far UV
(1500\AA) luminosity as our criterion to identify samples of
Lyman-break galaxies. We focus our attention on three redshifts,
$z=3$, $z=6$ and $z=10$, which are representative of the range covered
by current LBG studies \citep{Ste03,Bou07a,Bou10}.

\cite{Baugh05} presented predictions for the rest-frame UV luminosity
function at $z=3$.  Here we revisit this comparison, adding new
observational data at $z=3$ and going to $z=10$ (see also
\citealt{Lacey10b}).  In Fig.~\ref{LF} we show the predicted
rest-frame $1500$\AA\ luminosity function at $z=3$, $z=6$ and $z=10$,
including the effects of dust extinction, calculated in a
self-consistent way using the predicted scale lengths of the disk and
bulge components and the metallicity of the cold gas.  The model
matches the observational data at $z=3$, to within the scatter between
datasets, is in good agreement with the measurement by \cite{Bou07a}
at $z=6$, and is consistent with the tentative measurement and upper
limits estimated by \cite{Bou10} at $z=10$.  Following the common
practice in observational studies, we define a characteristic
luminosity $L^{*}_{UV}$ using the observed position of the break in
the UV luminosity function at $z=3$. We define a galaxy as a bright
LBG if its UV luminosity is brighter than $L^{*}_{UV}$ and as a faint
LBG if its UV luminosity exceeds $0.1 L^{*}_{UV}$ (a sample which
includes the bright LBG sample, but which is dominated in number by
galaxies close to the luminosity cut).  We apply this definition at
all redshifts. Note that $L^{*}_{UV}$ is taken to be the
characteristic luminosity at $z \approx 3$ found by \cite{Ste99}
(corresponding to $M^{*}_{UV}=-20.3+5{\rm log}h$ in our cosmology).
In Fig.~\ref{LF} we show where the $0.1 L^{*}_{UV}$ and $L^{*}_{UV}$
limits lie using vertical lines.

We note in passing that there is substantial evolution in both
the observed and predicted UV luminosity functions between $z=3$ and
$z=6$. The observed characteristic luminosity $L^{*}_{UV,z=6}$ at
$z=6$ ($M^{*}_{UV,z=6}=-19.5+5{\rm log}h$ in our cosmology,
\citealt{Bou07a}) is nearly a magnitude fainter than the $z=3$ value
of $L^{*}_{UV}$.  The number density of galaxies seen at the $z=3$
value of $L^{*}_{UV}$ drops by around a factor of $5$ between $z=3$
and $z=6$. At $z=10$, the abundance of galaxies with this luminosity
is predicted to be several hundred times lower than at $z=3$.  A
comprehensive study of the evolution of the LBG luminosity function,
including a more detailed comparison with observational data, is
presented in \cite{Lacey10b}.

\subsection{Example formation histories of model 
galaxies with Lyman-break progenitors}

As examples of the different galaxy formation and merger histories
which can produce bright LBGs, we plot in the left panels of
Fig.~\ref{GMT} galaxy merger trees for three present-day galaxies. The
mass of the galaxies increases down the page. The trees are
constructed by running {\tt GALFORM} with many output redshifts. We
stop plotting the trees at $z=6$ or when a branch falls below a
minimum galaxy mass (in stars and cold gas) of $10^{6}
h^{-1}M_{\odot}$.  A branch or trunk of the tree is plotted at each
output redshift using a circle. The size of the circle is proportional
to the stellar mass of the galaxy and the colour reflects the type of
galaxy: green for a faint LBG, red for a bright LBG and blue
otherwise.

The galaxy trees are plotted in the following way. Stepping back in
time from $z=0$, at each merger we plot the most massive branch on the
left and the other branches to the right of this.  The main progenitor
branch is hence the leftmost plotted branch traced back from $z=0$ in
this way.  Note that at $z>0$, the main progenitor branch does not
necessarily represent the most massive progenitor across the whole
tree at a given epoch.  The largest progenitor at any time could jump
from one branch of the galaxy tree to another, so its selection would
not necessarily define a smooth path back in redshift. We have chosen
to plot examples which have at least one LBG in the most massive
(leftmost) progenitor branch.

In the first example (Fig.~\ref{GMT} top left), we show the galaxy
merger tree for a galaxy which at $z=0$ has a stellar mass of
$M_{*}=2.5\times10^{9} h^{-1}M_{\odot}$ and has a bright LBG
progenitor at $z=6$.  In the second example (Fig.~\ref{GMT} middle
left), we show a galaxy with $M_{*}=6.6\times10^{10} h^{-1}M_{\odot}$
at $z=0$ with a bright LBG progenitor at both $z=6$ and $z=3$.  In the
third example (Fig.~\ref{GMT} bottom left), we show the galaxy merger
tree for a galaxy with $M_{*}=2.1\times10^{11} h^{-1}M_{\odot}$ at
$z=0$ with a bright LBG progenitor at $z=6$.  Note that these trees
are purely illustrative examples, chosen to show the range of
complexity of the trees, and are not intended to be a statistically
representative sample.

The right-hand panels of Fig.~\ref{GMT} show the evolution of various
properties of the galaxy in the most massive progenitor branch. The
upper inset or subpanel in each case shows the bulge-to-total stellar
mass ratio. If present, a vertical line in this panel marks the epoch
when the most massive progenitor ceased to be a central galaxy and
became a satellite. In the examples plotted here, this only happens in
the case of the least massive galaxy tree shown (upper right
Fig.~\ref{GMT}), at $z \sim 1.1$. The main right-hand panel shows the
UV luminosity of the most massive progenitor, in units of $L^{*}_{UV}$
(red line; see right axis for units) and its stellar mass (blue
symbols), cold gas mass (green symbols), stellar mass formed in bursts
(red symbols) and host dark halo mass (black symbols; see left axis
for units).  In the case of the least massive galaxy (Fig.~\ref{GMT}
top right), the most massive progenitor experiences a burst at $z=6$
which makes it a bright LBG. The stellar mass added to the bulge
during the burst increases the bulge to total stellar mass ratio, and
the cold gas supply is depleted.  Around $z \sim 5.2$ there is a
another burst of star formation, triggered by a major merger, which
leads to a large increase in the bulge-to-total ratio of the galaxy
from 0.3 to 0.9. However, there is less star formation associated with
this burst, due to the depletion of the gas reservoir prior to the
burst, and the most massive progenitor is only a faint LBG. Another
burst happens at $z \sim 3.5$, which makes the most massive progenitor
galaxy a faint LBG but with relatively little change in the $B/T$
ratio. The $B/T$ ratio gradually declines as quiescent star formation
adds mass to the disk. Finally, after a steady increase in mass, the
host halo is accreted by a much more massive halo at $z=1.1$, when the
most massive progenitor becomes a satellite galaxy. The present-day
galaxy is a satellite in a halo of mass $10^{14} h^{-1}M_{\odot}$. The
next most massive galaxy (middle panels Fig.~\ref{GMT}) displays two
epochs at which the most massive progenitor is a bright LBG ($z \sim
6$ and $z \sim 3.6$). The most massive progenitor branch in the most
massive galaxy (bottom panels of Fig.~\ref{GMT}) again shows several
instances when the progenitor is a bright LBG. It undergoes a major
merger at $z \approx 1.5$ which increases the stellar mass by 30\%
(barely visible in the figure), triggers a bright LBG phase, and
changes the morphology from $B/T \approx 0$ to $B/T \approx 1$.

It is interesting to note that in all cases shown, the stellar mass of
the most massive progenitor branch (shown by the blue points in the
right-hand panels) increases steadily with redshift, without
displaying any sudden large jumps. This implies that the bursts of
star formation in these examples are not responsible for large changes
in the stellar mass of the galaxy. This is a result in part of the large
recycled fraction ($R=0.91$) for stars forming with the top-heavy IMF,
together with the strong feedback from supernova explosions.

\subsection{The triggering and duration of the Lyman-break phase}
\label{sec:triggering}

\begin{figure}
\begin{center}
\includegraphics[width=8cm, bb= 85 360 560 730]{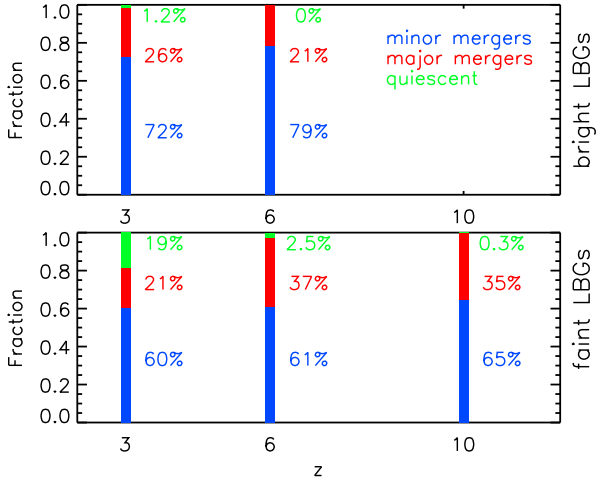}
\end{center}
\caption{The triggering mechanism of the LBG phase. For each model
LBG, we trace its history back in time from the redshift at which it
is identified, to determine whether the UV luminosity is dominated by
stars formed in a burst triggered by a minor merger (blue), a burst
triggered by a major merger (red) or stars formed quiescently
(green). The top panel shows the fractions for these different
triggering mechanisms for bright LBGs ($L_{UV} > L^{*}_{UV}$)
identified at $z=3$ and $6$, while the bottom panel shows the
fractions for faint LBGs ($L_{UV} > 0.1 L^{*}_{UV}$) identified at
$z=3$, $6$ and $10$. (No bright LBGs are predicted at $z=10$.)}
\label{fig:trigger}
\end{figure}

\begin{figure}
\begin{center}

\includegraphics[width=8cm, bb=85 360 550 710]{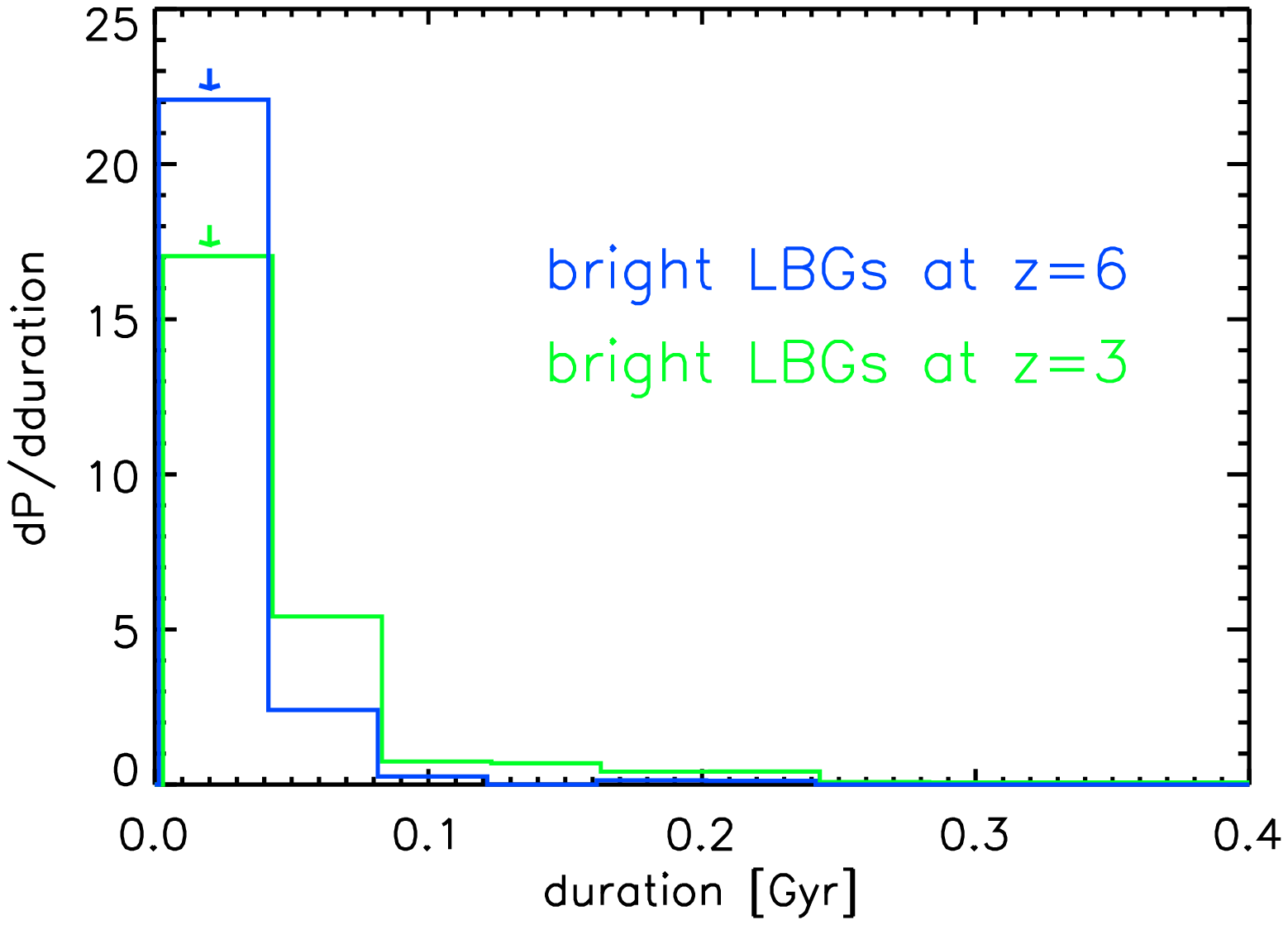}

\includegraphics[width=8cm, bb=85 360 550 710]{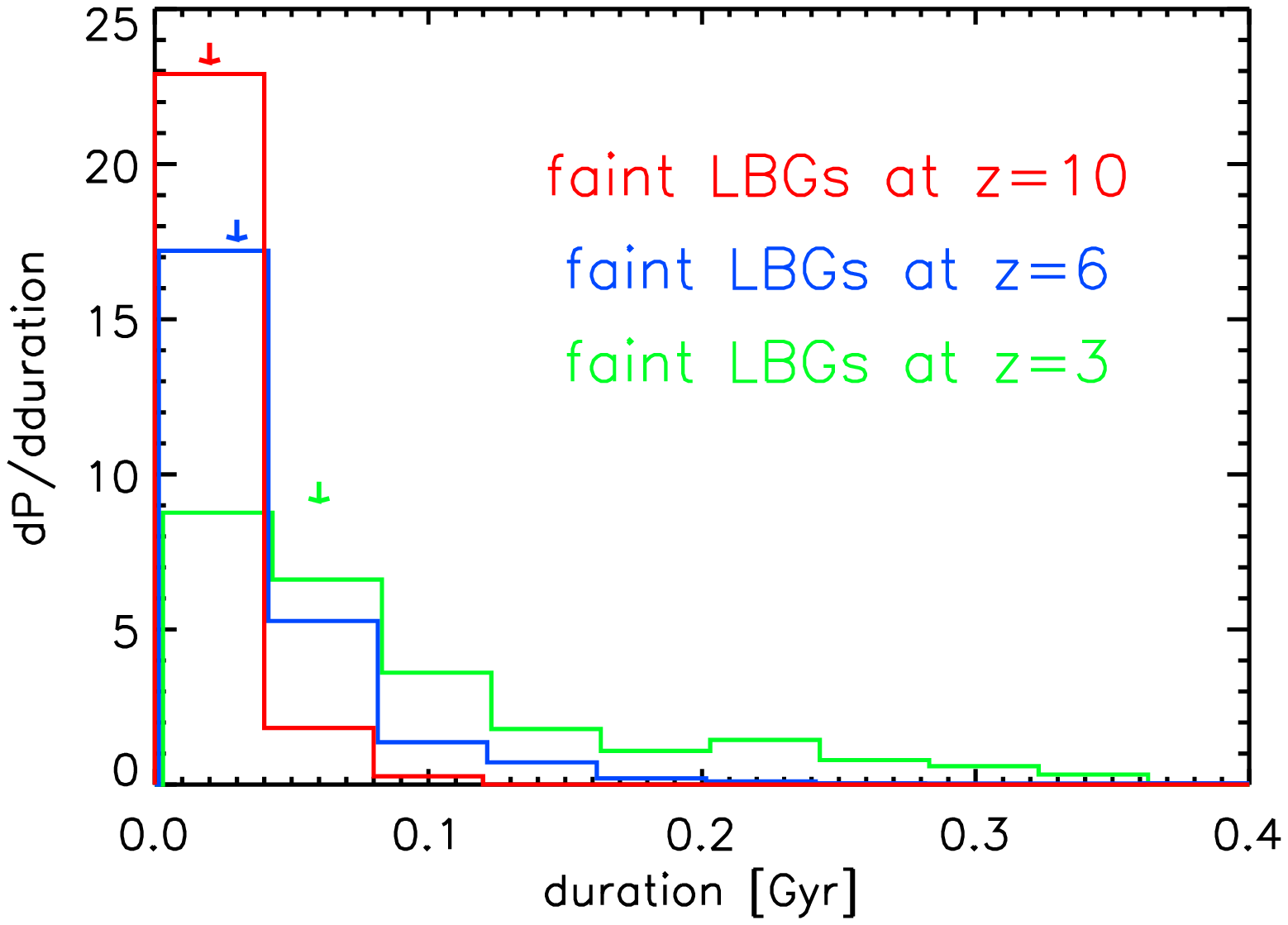}

\end{center}

\caption{The distribution of durations of the LBG phase. The top panel
shows the normalized distribution of durations for bright LBGs
($L_{UV} > L^{*}_{UV}$) identified at $z=3$ and $6$, while the bottom
panel shows this for faint LBGs ($L_{UV} > 0.1 L^{*}_{UV}$) identified
at $z=3$, $6$ and $10$. The vertical arrows show the median duration
for each LBG sample.}
\label{fig:duration}

\end{figure}

We can use the formation histories of model LBGs to answer some
important questions. The first of these is: what is the mechanism of
the star formation reponsible for the galaxy being seen as an LBG at a
particular redshift? In the present model, the possibilities are: a
starburst triggered by a minor galaxy merger; a starburst triggered by
a major galaxy merger; and quiescent star formation in a galaxy
disk. To determine which of these dominates in a particular case, we
move back in time through the galaxy formation history, starting at
the redshift at which a galaxy is identified as an LBG. If at any time
during the current LBG phase the burst and quiescent SFRs satisfy the
condition ${\rm SFR}_{\rm quiescent}> 4\, {\rm SFR}_{\rm burst}$ then
we classify the LBG phase as being produced by quiescent star
formation, otherwise we find the first galaxy merger (going back in
time) and identify whether it is a minor or major merger. We include
the factor 4 in the condition on the SFRs because stars forming with
the top-heavy burst IMF produce roughly 4 times higher UV luminosities
than the same mass of stars formed with the quiescent IMF. The results
of this classification exercise are shown in Fig.~\ref{fig:trigger},
for bright LBGs ($L_{UV} > L^{*}_{UV}$) identified at $z=3$ and $6$
(top panel), and for faint LBGs ($L_{UV} > 0.1 L^{*}_{UV}$) identified
at $z=3$, $6$ and $10$ (bottom panel). The fractions shown there are
calculated for a volume-limited sample of LBGs at the identification
redshift. We see from the figure that LBGs are predicted to be
dominated by starbursts triggered by minor galaxy mergers (around
60-80\% of cases) in all of the cases examined here. Minor mergers
dominate over major mergers by a factor 2--4. Quiescent star formation
is predicted to account for at most a few percent of LBGs, with the
exception of faint LBGs at $z=3$, where this fraction increases to
around 20\%. We also see that the quiescent fraction decreases with
increasing redshift for both bright and faint LBGs, and is much
smaller for bright LBGs than for faint LBGs at the same
redshift. These results about the dominance of bursts over quiescent
star formation in LBGs are in agreement with the analysis by
\citet{Lacey10b}.

A second question we can answer is: what is the time duration of the
LBG phase? For bright LBGs, we define this as the time during which
$L_{UV} > L^{*}_{UV}$ in the current LBG phase, while for faint LBGs
we define the duration as the time during which $L_{UV} > 0.1
L^{*}_{UV}$. We show the distribution of LBG durations for
volume-limited LBG samples in Fig.~\ref{fig:duration}, where the top
panel shows results for bright LBGs identified at $z=3$ and $6$, and
the lower panel faint LBGs identified at $z=3$, $6$ and $10$. For
bright LBGs, the median duration is around 20~Myr at both $z=3$ and
$z=6$. For faint LBGs, the durations are generally larger, as
expected, decreasing from around 60~Myr at $z=3$ to 30~Myr at $z=6$
and 20~Myr at $z=10$. The short durations for bright LBGs reflect the
relatively short timescales of the bursts responsible for most of
them.

\subsection[]{The fraction of present-day galaxies and halos with an LBG progenitor}

\begin{figure*}
\begin{center}

\includegraphics[width=14cm, bb=60 290 560 610]{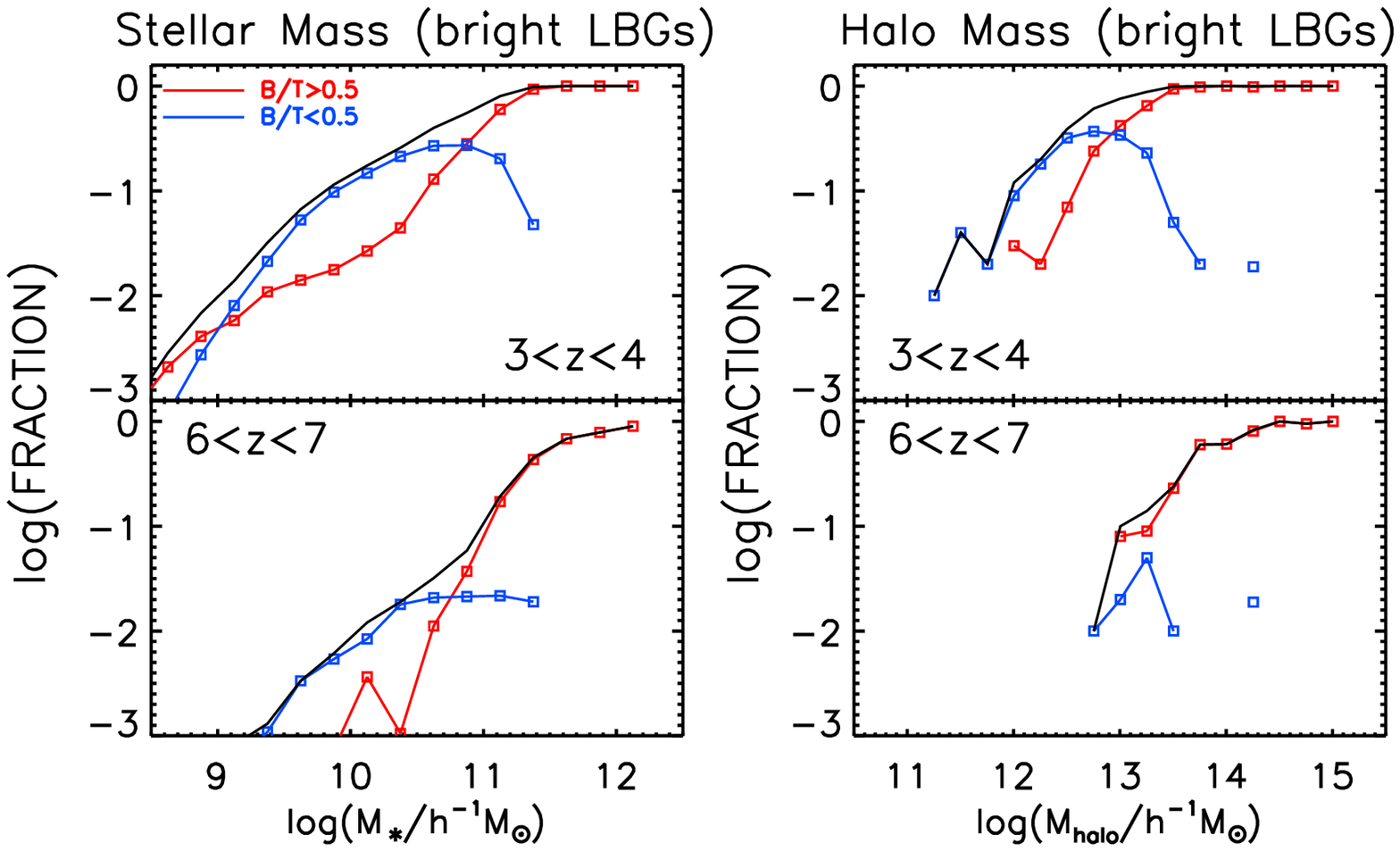}

\includegraphics[width=14cm, bb=60 170 560 610]{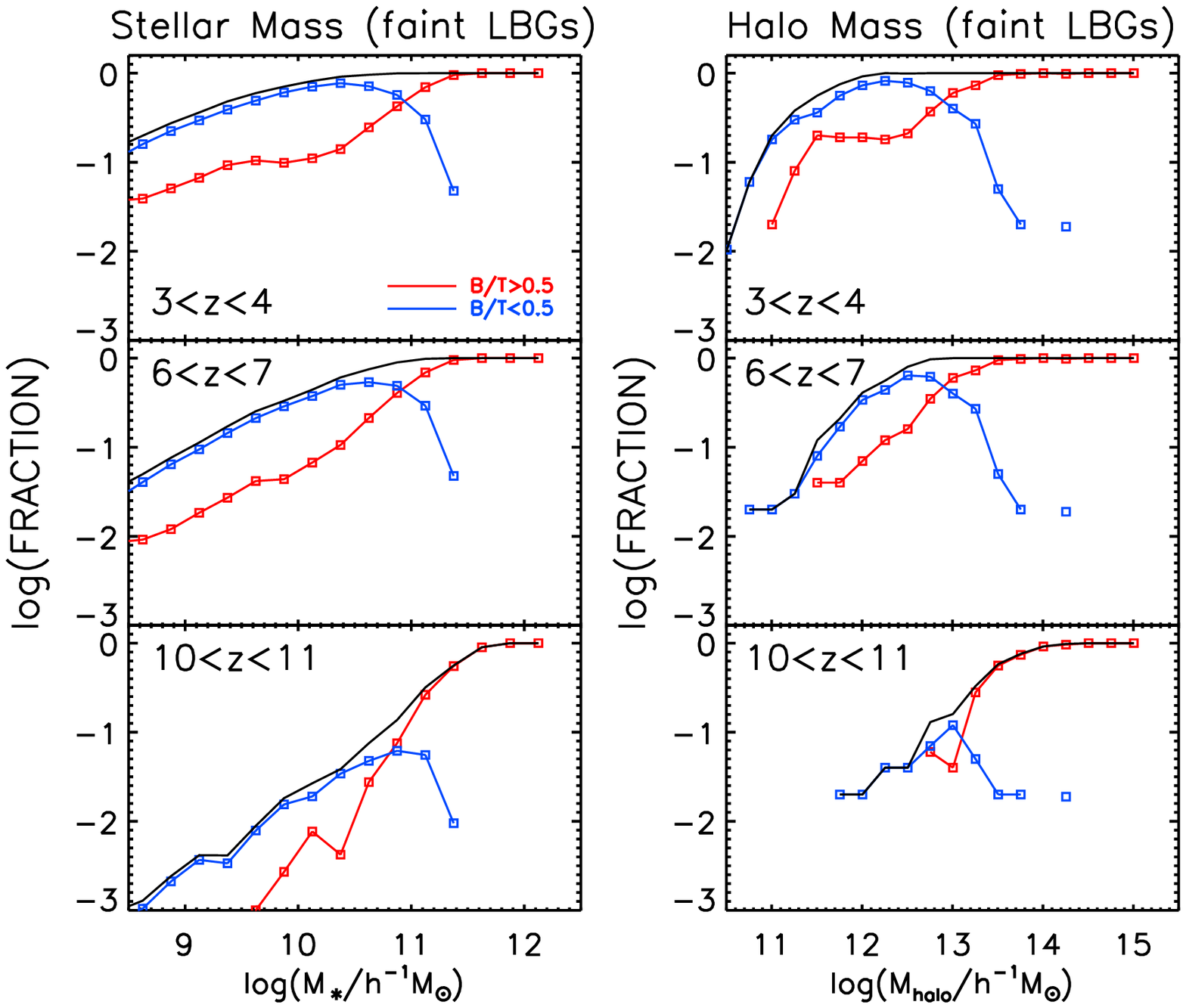}

\end{center}

\caption{ The predicted fraction of present day galaxies (left) and
haloes (right) with either bright LBG progenitors (i.e. $L_{UV} >
L^{*}_{UV}$) in top set of panels, or faint LBG progenitors
(i.e. $L_{UV} > 0.1 L^{*}_{UV}$) in bottom set of panels.  Left
panels: the fraction of present-day galaxies as a function of their
stellar mass which are predicted to have one or more LBG
progenitors. Right panels: the fraction of present-day halos as a
function of mass that are predicted to have hosted at least one LBG in
a progenitor halo. Fractions are shown for LBG progenitors at $3<z<4$,
$6<z<7$ and $10<z<11$ (the last for faint LBGs only), as
labelled. Descendant galaxies are separated into bulge-dominated
($B/T>0.5$, red) and disk-dominated ($B/T<0.5$, blue).  The black line
corresponds to the case of all LBG descendants and is the sum of the
blue and red lines.  }
\label{FRACTIONLBG}

\end{figure*}

Having shown some illustrative examples of galaxy merger trees with
LBG progenitors in the previous subsection, we now perform a
statistical study and compute the probability that a present-day
galaxy had a progenitor that was an LBG or that a present-day halo had
a progenitor halo which hosted an LBG.

The predictions for the fraction of present-day galaxies that are
descendants of LBGs are presented in Fig.~\ref{FRACTIONLBG}, with the
top panels showing results for descendants of bright LBGs at $3<z<4$
and $6<z<7$, and the lower panels results for descendants of faint
LBGs at $3<z<4$, $6<z<7$ and $10<z<11$. (The model does not predict a
significant number bright LBGs at $z=10$.) In the left-hand panels we
plot the fraction of present-day galaxies, as a function of their
stellar mass, which are predicted to have at least one bright LBG
progenitor in the specified redshift range. In the right-hand panels
we show the fraction of present-day halos having progenitor halos
which hosted an LBG in the specified redshift range. We see that the
probability to have an LBG progenitor increases with increasing
present-day mass (both stellar and halo), but decreases with
increasing LBG redshift

Considering first the bright LBGs, we find that galaxies with stellar
masses of $6.3\times 10^{9} h^{-1}M_{\odot}$, $6.1\times 10^{10}
h^{-1}M_{\odot}$ and $1.8\times 10^{11} h^{-1}M_{\odot}$ are predicted
to respectively have 10\%, 50\% and 90\% probabilities of having at
least one bright LBG progenitor in the range $3<z<4$. The
corresponding stellar masses for bright LBGs at $6<z<7$ are 4-15 times
larger, being $9.0\times 10^{10} h^{-1}M_{\odot}$, $2.7\times 10^{11}
h^{-1}M_{\odot}$ and $1.3\times 10^{12} h^{-1}M_{\odot}$ respectively.
A Milky Way mass galaxy ($4\times10^{10} h^{-1}M_{\odot}$,
\citealt {BinneyGalacticDynamics}) is predicted to have a 40\% or 3\%
probability of having a bright LBG progenitor at $3<z<4$ or $6<z<7$,
respectively.  The right hand panels of Fig.~\ref{FRACTIONLBG} show
that 50\% of present-day haloes with a mass greater than $4.3\times
10^{12} h^{-1}M_{\odot}$ are predicted to have halo progenitors that
hosted bright LBGs at $3<z<4$. The corresponding mass for bright LBGs
at $6<z<7$ is an order of magnitude larger, $4.8\times 10^{13}
h^{-1}M_{\odot}$.

Considering now the faint LBGs, we find that 10\%, 50\% and 90\% of
galaxies with present-day stellar masses of $1.3\times 10^{8}
h^{-1}M_{\odot}$, $2.6\times 10^{9} h^{-1}M_{\odot}$ and $2.2\times
10^{10} h^{-1}M_{\odot}$ are predicted to have had faint LBG
progenitors at $3<z<4$. These masses increase steeply to $1.1\times
10^{9} h^{-1}M_{\odot}$, $1.6\times 10^{10} h^{-1}M_{\odot}$ and
$7.7\times 10^{10} h^{-1}M_{\odot}$ for faint LBGs at $6<z<7$, and
steeply again to $5.3\times 10^{10} h^{-1}M_{\odot}$, $2.0\times
10^{11} h^{-1}M_{\odot}$ and $4.2\times 10^{11} h^{-1}M_{\odot}$ at
$10<z<11$.  A Milky Way mass galaxy is predicted to have a 95\% or
70\% probability of having a faint LBG progenitor at $3<z<4$ or
$6<z<7$ respectively. This probability drops to 7\% for a $10<z<11$
faint LBG progenitor.  Expressed in terms of halo masses (right-hand
panels of Fig.~\ref{FRACTIONLBG}), we find that 50\% of halos today
with masses larger than $2.6\times10^{11} h^{-1}M_{\odot}$ are
predicted to have halo progenitors that hosted faint LBGs at $3<z<4$,
while for $6<z<7$ this increases to $1.4\times10^{12}
h^{-1}M_{\odot}$. Only for masses greater than $2.7 \times 10^{13}
h^{-1}M_{\odot}$ are 50\% of haloes predicted to have had at least one
faint LBG progenitor at $10<z<11$.

\section{The properties of LBGs and their descendants}

Having defined samples of bright and faint LBGs, we now compare the
model predictions for their stellar mass, host halo mass and
morphologies at redshifts $z=3$, $z=6$ and $z=10$. The semi-analytical
model allows us to follow the subsequent evolution of the LBGs, so we
also present the predictions for the analogous properties of the
descendants of the LBGs. We also consider what is the dominant mode of
star formation - burst or quiescent - in LBGs. Note that at $z=10$,
the model does not predict a significant population of bright LBGs
(see Fig.~\ref{LF}). When we construct the distribution of properties
of the descendants of LBGs, each descendant galaxy is counted once
(with a weight depending on its abundance), regardless of the number
of LBG progenitors it possesses.

\subsection{Stellar masses}
\label{sec:stellar_mass}

\begin{figure*}
\begin{center}

\includegraphics[width=14cm, bb=60 290 560 610]{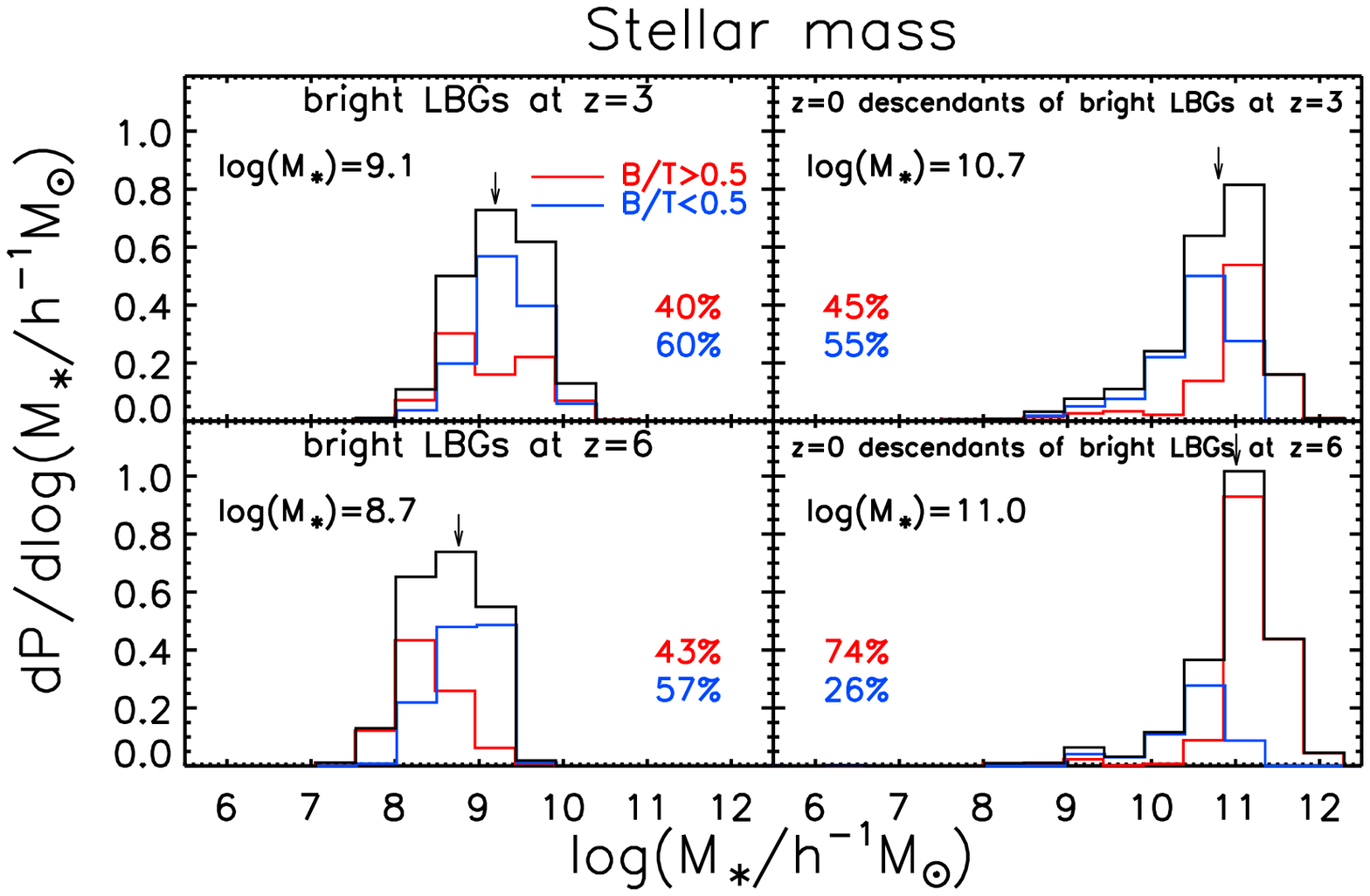}

\includegraphics[width=14cm, bb=60 170 560 610]{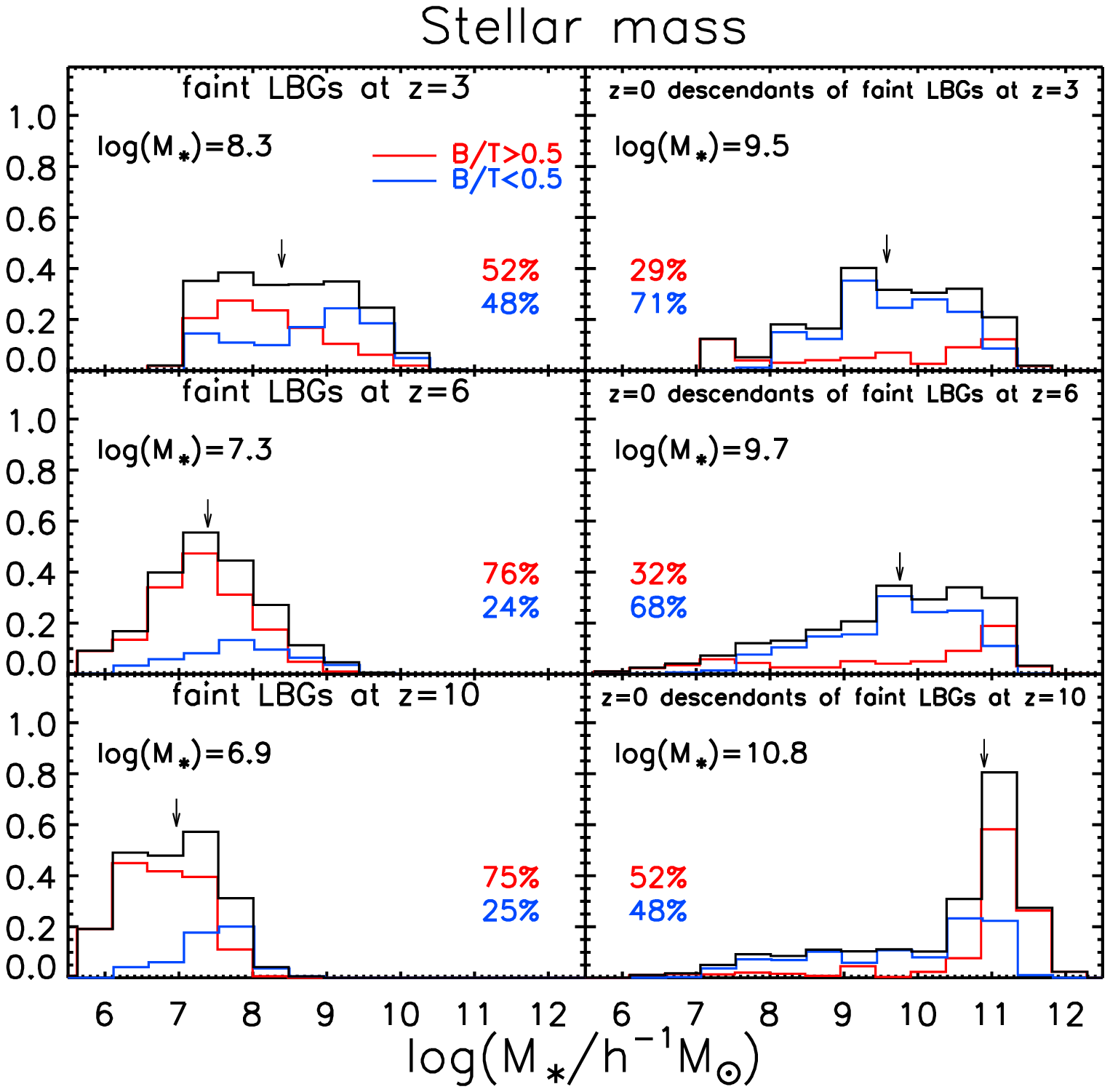}

\end{center}

\caption{ The stellar mass distribution of LBGs (left) and their z=0
descendants (right). The top set of panels show predictions for bright
LBGs ($L_{UV} > L^{*}_{UV}$), and the bottom set of panels faint LBGs
($L_{UV} > 0.1 L^{*}_{UV}$). Results are shown for $z=3$, $z=6$ and
$z=10$ (the last for faint LBGs only) as indicated in each panel. The
median stellar mass value is given in the label and is indicated by an
arrow. Galaxies are separated into bulge-dominated ($B/T>0.5$, red
histogram) and disk-dominated galaxies ($B/T<0.5$, blue
histogram). The percentage of galaxies in each class is labelled in
the same colour as the corresponding histogram. The red and blue
distributions add to give the total stellar mass distribution (black
histogram).  }
\label{mstdesc}

\end{figure*}

The model predicts that the stellar masses of LBGs identified at a
given UV luminosity decrease with increasing redshift. On the other
hand, the stellar masses of their present-day descendants show the
opposite trend, increasing with increasing redshift. This behaviour is
seen for both bright and faint LBGs, as we now discuss in detail.

We plot the stellar mass distributions of LBGs and of their
present-day descendants in the left and right panels of
Fig.~\ref{mstdesc}. The top panels show results for bright LBGs
($L_{UV} > L^{*}_{UV}$) at $z=3$ and $z=6$, while the lower panels
show results for faint LBGs ($L_{UV} > 0.1 L^{*}_{UV}$) at $z=3$,
$z=6$ and $z=10$.

We see that the median stellar mass of bright LBGs at $z=3$ is
$M_{*}=1.3\times10^{9} h^{-1}M_{\odot}$, which is 6 times larger than
that of their faint counterparts at this redshift, and $\approx 2.5$
times larger than the median mass of bright LBGs at $z=6$. The median
stellar mass of the $z=0$ descendants (right panels) of these two LBG
populations ($z=3$ and $z=6$) are predicted to be
$M_{*}=5\times10^{10} h^{-1}M_{\odot}$ for the bright LBGs at $z=3$
and $M_{*}=1\times 10^{11} h^{-1}M_{\odot}$ for the bright LBGs at
$z=6$. We separate the galaxies into bulge-dominated (if the bulge to
total stellar mass ratio $B/T > 0.5$) and disk-dominated galaxies
(i.e. $B/T < 0.5$).  Bright LBGs at $z=3$ and $z=6$ are predicted to
be typically (60\%) disk-dominated. However, while the descendants of
bright LBGs at $z=3$ are also mainly disk-dominated, descendants of
bright LBGs at $z=6$ are mainly bulge-dominated.

The median stellar mass of faint LBGs at $z=3$ is $M_{*}=2\times10^{8}
h^{-1}M_{\odot}$, which is an order of magnitude larger than the
median mass of faint LBGs at $z=6$. Faint LBGs at $z=10$ are 25 times
less massive than faint LBGs at $z=3$. The median stellar masses of
the descendants of the $z=3$ and $z=6$ populations are predicted to be
similar ($M_{*} \sim (3-5)\times10^{9} h^{-1}M_{\odot}$). The median
descendant mass of a $z=10$ faint LBG is an order of magnitude larger
than for faint LBGs identified at $z=3$ or $z=6$.  The descendants of
LBGs show a broader distribution of stellar mass compared with the
LBGs themselves.  We also find that faint LBGs at $z=6$ and $z=10$ are
predominantly bulge-dominated systems (75\% are predicted to have
$B/T>0.5$).  At $z=3$ there is a more balanced population, with almost
half of the faint LBGs predicted to be disk-dominated systems.  The
descendants of faint LBGs identified at $z=3$ and $z=6$ have similar
morphological mixes. They are predicted to be mainly disk-dominated
systems (with 70\% having $B/T<0.5$). The descandants of $z=10$ faint
LBGs are equally likely to be disk or bulge dominated.  Among the
descendants, those of lower mass ($M \lesssim 10^{11}
h^{-1}M_{\odot}$) are typically predicted to be disk-dominated
systems, whereas higher mass LBG descendants ($M \gsim
10^{11}h^{-1}M_{\odot}$) are mostly bulge-dominated systems in the
model.

Stellar masses of LBGs have been estimated photometrically in a number
of observational studies
\citep[e.g][]{Papovich01,Shapley01,Verma07,Stark09}. These are all
based on fitting model SEDs to multicolour photometry.  As discussed
in \citet{Lacey10b}, these photometric mass estimates are fraught with
difficulty, since they depend sensitively on the assumed IMF, and also
on assumptions about dust extinction, star formation history and
metallicity. Since our model galaxies contain a mixture of stars
formed with quiescent and burst IMFs, while the observational
estimates all assume a solar neighbourhood (typically Salpeter) IMF, a
direct comparison between stellar masses predicted by the model and
values estimated observationally from photometric data from is not
meaningful. Instead, \citeauthor{Lacey10b} made a direct photometric
comparison between the models and observations. They plot the
observer-frame flux at $3.6 \mu {\rm m}$ against rest-frame far-UV
flux, and find, using the same model as we do here, that the
predictions come to within a factor of 3 of matching the observed
relation found by \citet{Stark09} for LBGs at $z=4-6$. At the
redshifts of the LBG samples, the $3.6 \mu {\rm m}$ flux probes the
optical rest-frame, which is determined by the recent star formation
history of a galaxy as well as its stellar mass. We will investigate
this issue in more detail in a future paper.

\subsection{Host halo masses}
\label{sec:halo_mass}

\begin{figure*}
\begin{center}

\includegraphics[width=14cm, bb=60 290 560 610]{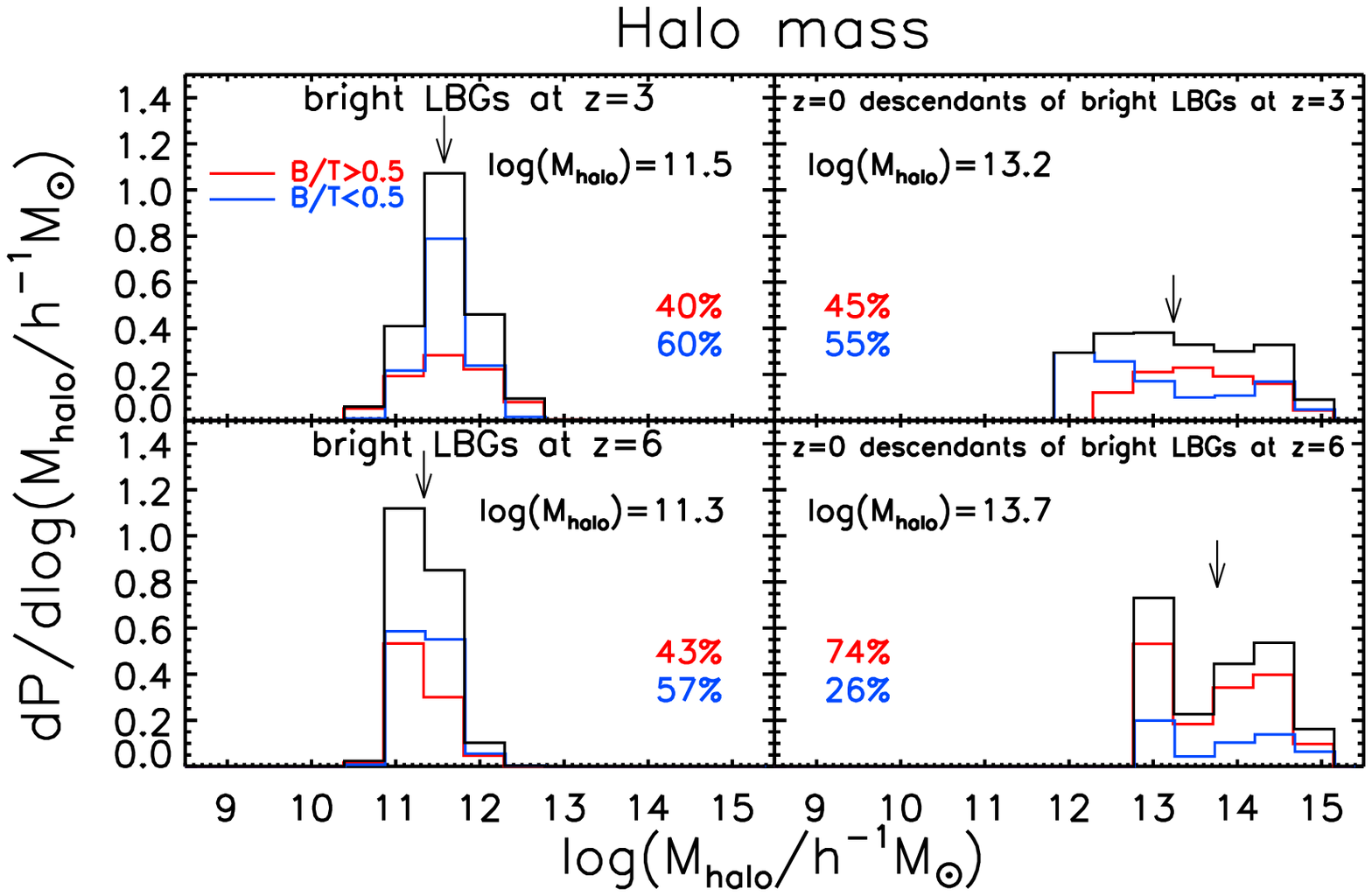}

\includegraphics[width=14cm, bb=60 170 560 610]{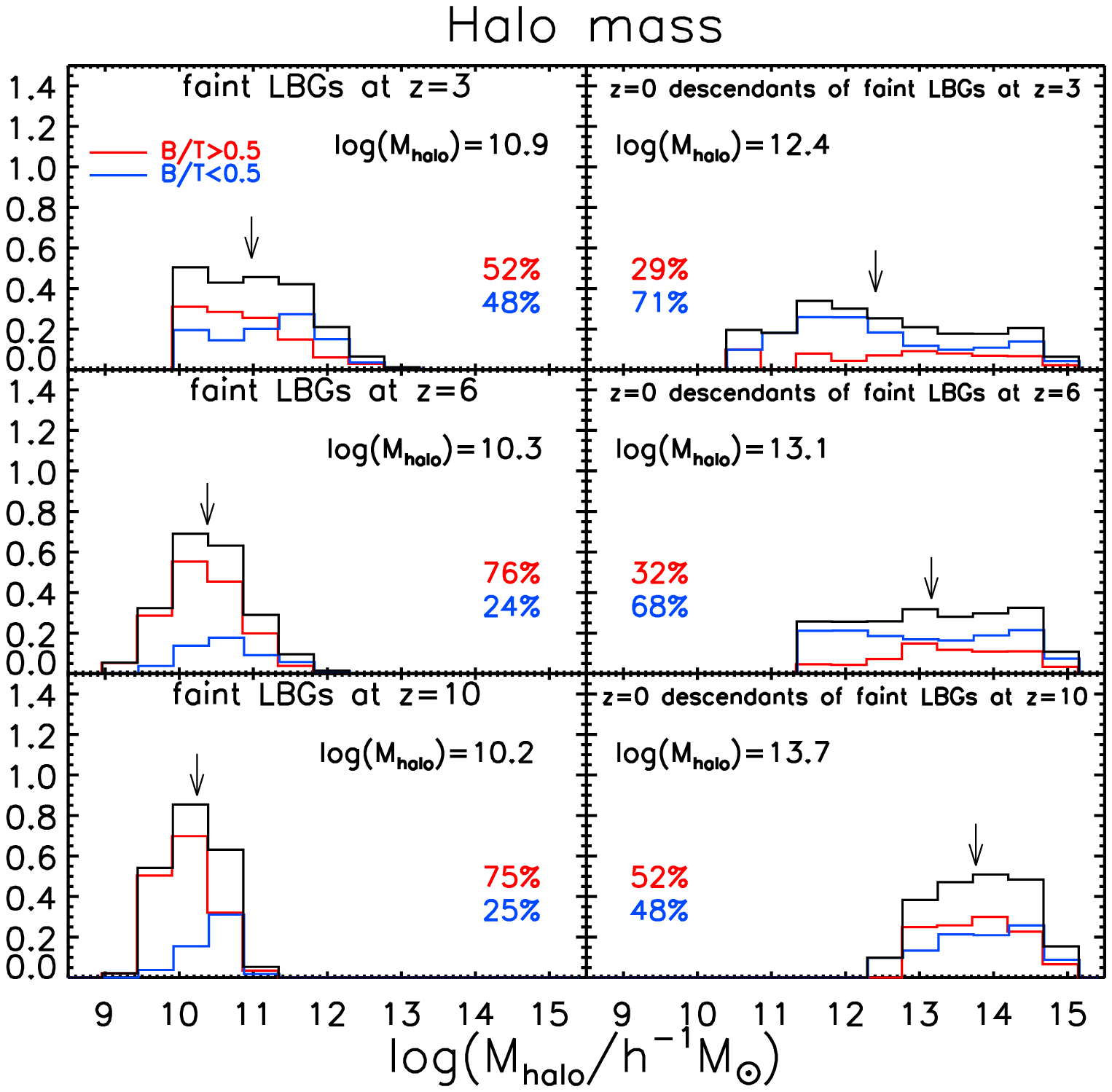}

\end{center}

\caption{ Host halo mass distribution of LBGs (left) and the host halo
mass distribution of their $z=0$ descendants (right). The top set of
panels show bright LBGs ($L_{UV} > L^{*}_{UV}$) and the bottom set of
panels show faint LBGs ($L_{UV} > 0.1 L^{*}_{UV}$). Predictions are
shown for $z=3$, $z=6$ and $z=10$ (the last for faint LBGs only), as
indicated in each panel.  The median host halo mass value is written
in each panel and indicated by an arrow.  The red and blue histograms
shows the predictions for bulge and disk dominated galaxies
respectively; the percentage of galaxies in each class is given in the
legend, using the same colour.  }
\label{mhalodesc}

\end{figure*}

The model predicts that the host halo masses of LBGs and of their
present-day descendants show trends with LBG redshift in the same
sense as for the stellar masses, i.e. host halo masses for LBGs of a
given UV luminosity decrease with increasing redshift, while the host
halo masses of their descendants present-day increase. We now discuss
this in detail for bright and faint LBGs.

We plot the predicted distributions of host halos masses of LBGs and
of their present-day descendants in the left and right panels of
Fig.~\ref{mhalodesc}. The upper panels show results for bright LBGs at
$z=3$ and $z=6$, while the lower panels show results for faint LBGs at
$z=3$, $z=6$ and $z=10$.

We see that bright LBGs at $z=3$ live in halos with a median mass of
$3\times10^{11} h^{-1}M_{\odot}$, while bright LBGs at $z=6$ reside in
halos with a median mass of $2\times10^{11} h^{-1}M_{\odot}$.  The
descendants of bright LBGs identified at $z=3$ are hosted by halos
with a median mass of $1.6 \times 10^{13} h^{-1}M_{\odot}$, and the
$z=0$ descendants of bright $z=6$ LBGs are hosted by halos with a
median mass of $5\times10^{13} h^{-1}M_{\odot}$.

Faint LBGs at $z=3$ live in halos with a median mass of
$8\times10^{10} h^{-1}M_{\odot}$, which is 4 times larger than the
halos which host faint LBGs at $z=6$. The distribution of host halo
mass for faint LBGs at $z=10$ is similar to that at $z=6$.  The
descendants of faint $z=3$ LBGs are hosted by halos with a median mass
of $2.5\times10^{12}h^{-1}M_{\odot}$, and the descendants of faint
$z=6$ LBGs are hosted by halos with a median mass of $1.3 \times
10^{13} h^{-1}M_{\odot}$. Halos which host the descendants of faint
LBGs have a broader distribution in mass than halos which host the
progenitors of bright LBGs.

The host halo masses of LBGs can be observationally constrained
through measurements of their clustering. \citet{Lacey10b} show
predictions for the linear clustering bias $b$ of LBGs as a function of
luminosity and redshift for the same model as used here. For bright
LBGs, they find $b\approx 2$ at $z=3$, increasing to $b \approx 4$ at
$z=6$. As they discuss, these predictions seem broadly consistent with
observational estimates of the clustering bias by \citet{Adelberger98}
and \citet{Ouchi04}.

Earlier calculations with a different semi-analytical model suggested
that LBGs reside in more massive haloes ($\approx 2 \times 10^{12}
h^{-1} M_{\odot}$) at $z=3$ \citep{Baugh98,Governato98}. However, this
earlier model had several important differences from the
\citet{Baugh05} model used in the present paper. It did not include
the effect of dust extinction on the UV luminosity of the model
galaxies, it invoked a much stronger supernova feedback in low
circular velocity haloes, it had much less star formation in
merger-driven starbursts at high redshift, and it did not assume a
top-heavy IMF in bursts. The last three of these differences
mean that in the \citet{Baugh05} model, objects can be bright in the
UV (and thus be classified as LBGs) while having lower masses than in
the \citet{Baugh98} model.


\subsection{Morphology}

\begin{figure*}
\includegraphics[width=14cm]{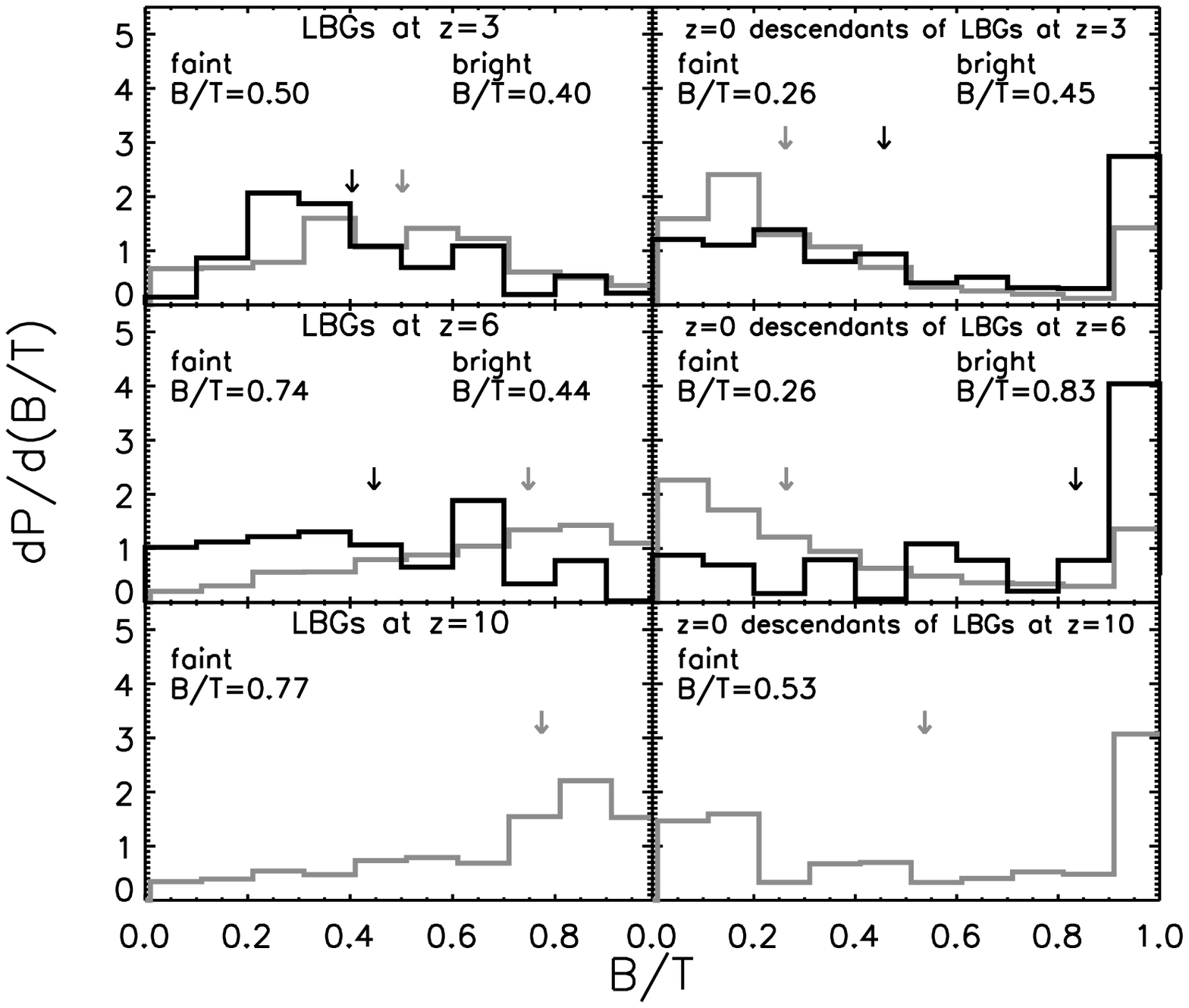}
\caption{ The distribution of bulge to total stellar mass ratio (B/T)
for LBGs (left) and their $z=0$ descendants (right). The distributions
for faint LBGs ($L_{UV} > 0.1 L^{*}_{UV}$) are plotted in grey and for
bright LBGs ($L_{UV} > L^{*}_{UV}$) in black. The top row shows
predictions for LBGs identified at $z=3$ and their descendants.  The
middle row shows the equivalent predictions for $z=6$ LBGs and the
bottom row for $z=10$ LBGs.  The median $B/T$ value is written in each
panel and indicated by an arrow in the same colour as the
corresponding histogram.  }
\label{BTdesc}
\end{figure*}

In \S\ref{sec:stellar_mass} we showed that the model predicts a
mixture of morphological types (as defined using the bulge to total
stellar mass ratio, $B/T$) for LBGs and their descendants.
Fig.~\ref{BTdesc} shows the distribution of bulge-to-total stellar
mass ratio ($B/T$) for bright and faint LBGs at $z=3$, $z=6$ and
$z=10$, and for their present-day descendants. We see that faint LBGs
are typically more bulge-dominated in stellar mass than bright LBGs at
the same redshift. However, for their descendants, this trend is
reversed, with the descendants of bright LBGs being more
bulge-dominated. There is also a tendency for LBGs at a given
luminosity (whether bright or faint) to become more bulge-dominated
with increasing redshift, and the same trend is also seen for their
present-day descendants.

The observed morphologies of LBGs have mostly been studied by imaging
in their rest-frame UV light.  \citet{Lacey10b} present predictions
for the $B/T$ ratios of LBGs as seen in rest-frame UV luminosity for
the same model as used here, and compare with the observational
studies by \citet{Ravindranath06} and \citet{Lotz06}. These studies
find that for bright LBGs at $z\sim 3-5$, $\sim 30\%$ are bulge-like,
$\sim 10-30\%$ are major mergers, and $\sim 40-60\%$ are disk-like or
minor mergers. \citeauthor{Lacey10b} conclude that the model
predictions are qualitatively consistent with these observational
results.

\subsection{Mode of star formation in LBGs}

\begin{figure*}
 \includegraphics[width=14cm,bb=50 400 600 730]{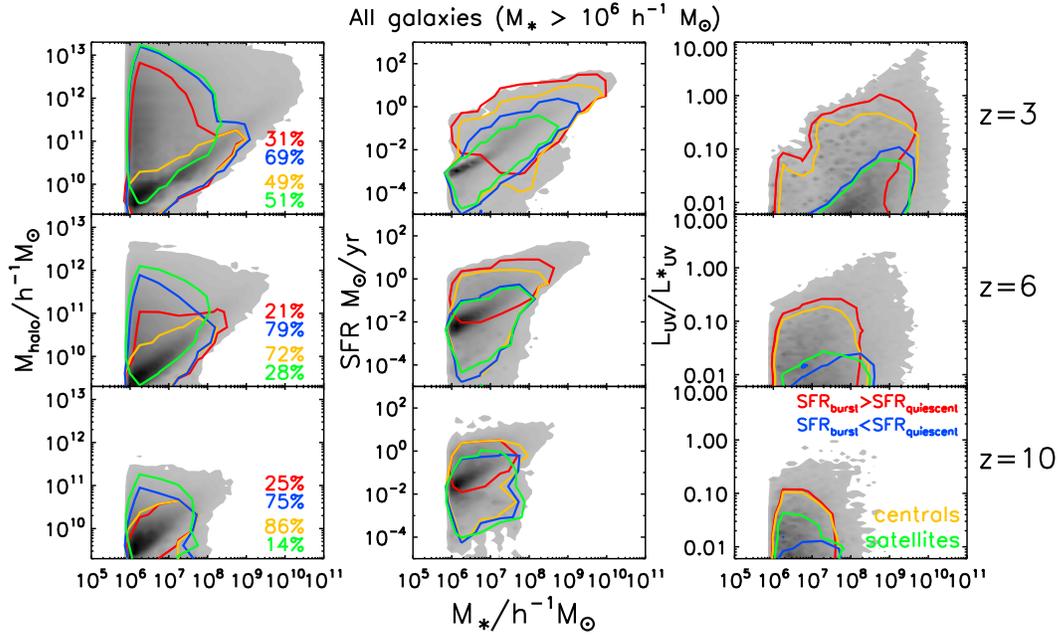}
\caption{ Various galaxy properties plotted against stellar mass for
$z=3$ (upper panels), $z=6$ (middle panels) and $z=10$ (lower panels):
on the $y$-axis in the left-hand panels is the host halo mass, in the
middle panels, the total star formation rate and in the right-hand
panels, the UV luminosity in units of $L^{*}_{UV}$ (defined at $z=3$).
The grey shading reflects the square root of the space density of
objects. 
Only galaxies with a stellar mass in excess of $10^{6}
h^{-1}M_{\odot}$ are included.  The coloured contours enclose 95\% of
the total space density when an additional selection is applied. In
the case of the red and blue contours, the extra information is the
dominant mode of star formation; the red contour encloses 95\% of the
galaxies for which the star formation in bursts exceeds the quiescent
star formation rate and the blue contour shows the converse. The
orange contour marks the locus within which 95\% of central galaxies
are to be found and the green contour shows the corresponding
distribution of satellites. The labels of the same colour as the
contours give the percentage of galaxies within each class.
  }
\label{MvsMSFRALL}
\end{figure*}

\begin{figure*}
 \includegraphics[width=14cm,bb=50 400 600 730]{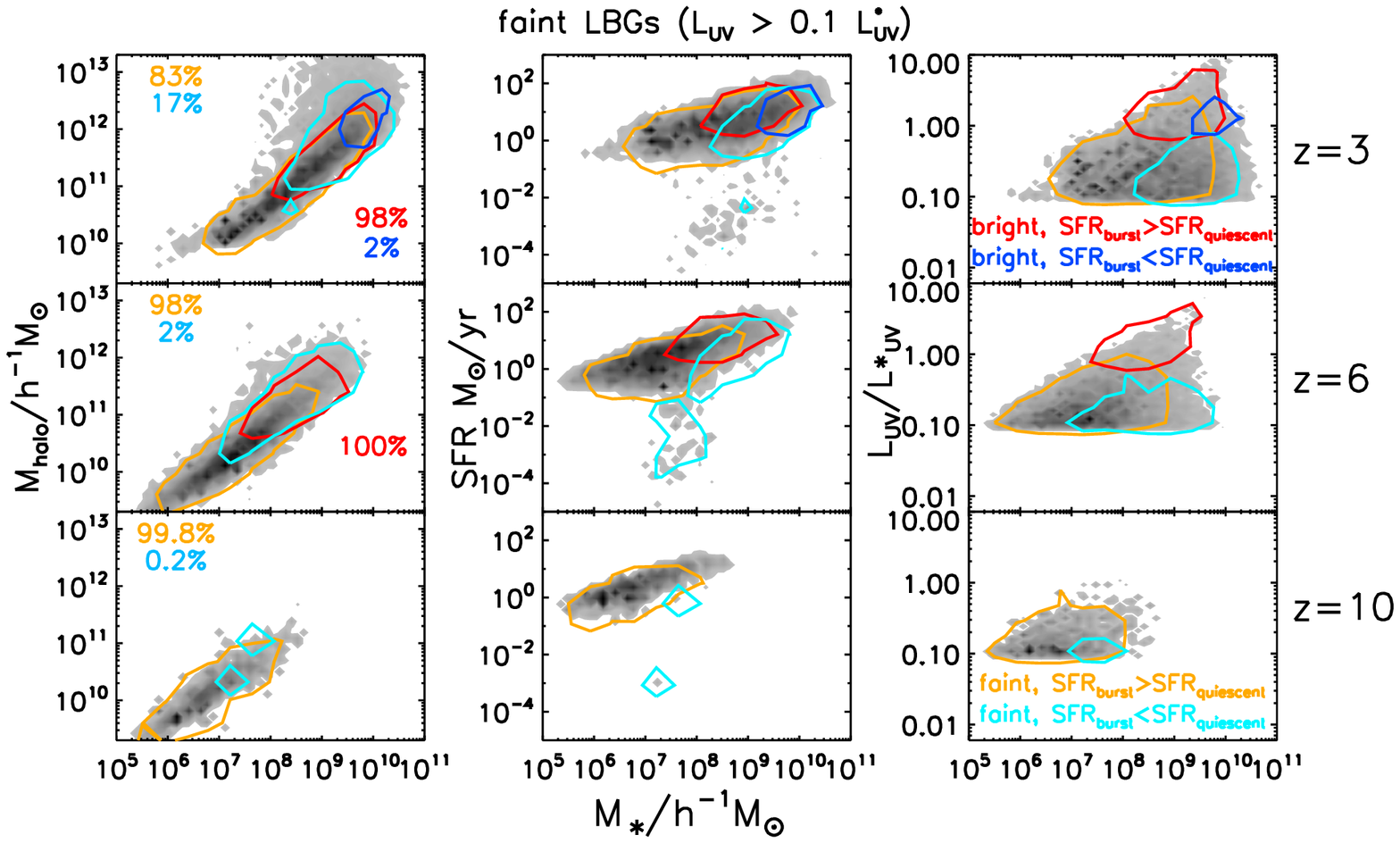}
\caption{ Similar to Fig.~\ref{MvsMSFRALL}, but restricting attention
to the properties of faint LBGs (i.e. $L_{UV}/L^{*}_{UV}>0.1$), as can
be seen clearly from the right-hand panels.  
As before, the coloured contours mark the locus of property space
which encloses 95\% of the selected sample. Galaxies in which the star
formation rate in bursts exceeds the quiescent star formation rate are
bounded by red contours for bright LBGs and orange contours for faint
LBGs; the galaxies in which quiescent star formation is dominant are
shown by the blue contours for bright LBGs and cyan contours for faint
LBGs. The percentages written in the same colour as the contours show
the overall proportions of galaxies in each class.
}
\label{MvsMSFRLBGs}
\end{figure*}

As discussed in Section 2, the model assumes two modes of star
formation, quiescent star formation in disks and star formation in
bursts triggered by galaxy mergers.  Quiescent star formation produces
stars with a standard solar neighbourhood IMF, whereas in burst star
formation, a top heavy IMF is adopted.  This difference in the choice
of IMF between the two modes of star formation has an impact on the
abundance and properties of LBGs. In this section we study
correlations between the star formation rate, stellar mass, halo mass
and UV luminosity of galaxies at $z=3$, $z=6$ and $z=10$. Our goal is
to understand the role of the different star formation channels in the
production of LBGs and in determining their nature.

In Fig.~\ref{MvsMSFRALL}, we plot all galaxies with stellar masses
$M_{*} > 10^{6} \hMsol$ at $z=3, \,6$ and $10$, showing the relations
between stellar mass (plotted on the $x$-axis) and various properties
on the $y$-axis: host halo mass (left panel), star formation rate
(middle panel) and far-UV (1500\AA) luminosity in units of
$L^{*}_{UV}$ (right panel).  (We choose a lower mass limit of $M_{*} =
10^{6} \hMsol$ because the masses of faint LBGs at $z=10$ extend down
to roughly this mass.) The shading in Fig.~\ref{MvsMSFRALL} reflects
the square root of the space density of galaxies, becoming darker
where more galaxies are found.  We gain further insight by dividing
the galaxies according to which mechanism of star formation is
dominant (but without the factor 4 used earlier in
\S\ref{sec:triggering}) and whether the galaxies are central or
satellites in their host dark halo.  Contours of a given colour
enclose 95\% of the galaxies passing one of these selection criteria.
The red contours delimit the region occupied by 95\% of the galaxies
in which starbursts are the dominant mode of star formation and blue
contours enclose the quiescent population. Similarly, orange contours
show where the majority of central galaxies lie and green contours
show where the satellite galaxies fall.
 
Fig.~\ref{MvsMSFRALL} shows that there is a correlation between galaxy
stellar mass and host halo mass which is driven by central galaxies.
Satellite galaxies in the model, which are mostly quiescent, are
responsible for the large scatter in this relation (as suggested by
the green contour).  There is also a correlation between stellar mass
and star formation rate.  In this case the correlation is largely due
to galaxies where quiescent star formation dominates (as suggested by
the blue contours), with the scatter produced mainly by galaxies which
are undergoing bursts of star formation (see also \citealt{Lagos10}).

In Fig.~\ref{MvsMSFRLBGs} we plot similar correlations, but in this
case for faint LBGs ($L > 0.1 L^{*}_{UV}$) only.  Again, we
differentiate between galaxies according to the dominant mode of star
formation, and make this distinction for the faint and bright LBG
samples.  Fig.~\ref{MvsMSFRLBGs} shows that most LBGs are produced by
bursts of star formation, as discussed earlier in
\S\ref{sec:triggering}, and also by \citet{Lacey10b}. We again see
that the fraction of LBGs dominated by burst star formation increases
with redshift. The correlations of halo mass and SFR with stellar mass
are seen to be better defined, with less scatter, for LBGs than for
samples containing all galaxies. However, the correlation of SFR with
stellar mass is also seen to be different (shifted to higher SFRs) for
LBGs from that for all galaxies. This is because the LBG samples are
dominated by bursts, which have larger SFRs than quiescent galaxies of
the same stellar mass.


We have also investigated what fraction of the stellar mass of model
LBGs has formed quiescently or in bursts (allowing for the recycling
of mass from dying stars). We find that LBGs at $z=3$ typically formed
most of their stellar mass by quiescent star formation, with 57\% of
faint LBGs and 73\% of bright LBGs having formed more than half of
their stellar mass quiescently. However, at higher redshifts, the
reverse is true: at $z=6$, 93\% of faint LBGs and 89\% of bright LBGs
have built at least half their stellar mass through bursts of star
formation after merging with other galaxies, while at $z=10$, 99\% of
faint LBGs have built at least half their stellar mass through
bursts. (See also the discussion in \citealt{Lacey10b}.)


\section{What is the connection between LBGs and SMGs?}

\begin{figure*}
\begin{center}

\begin{minipage}{8cm}
\includegraphics[width=8cm, bb=55 35 550 760]{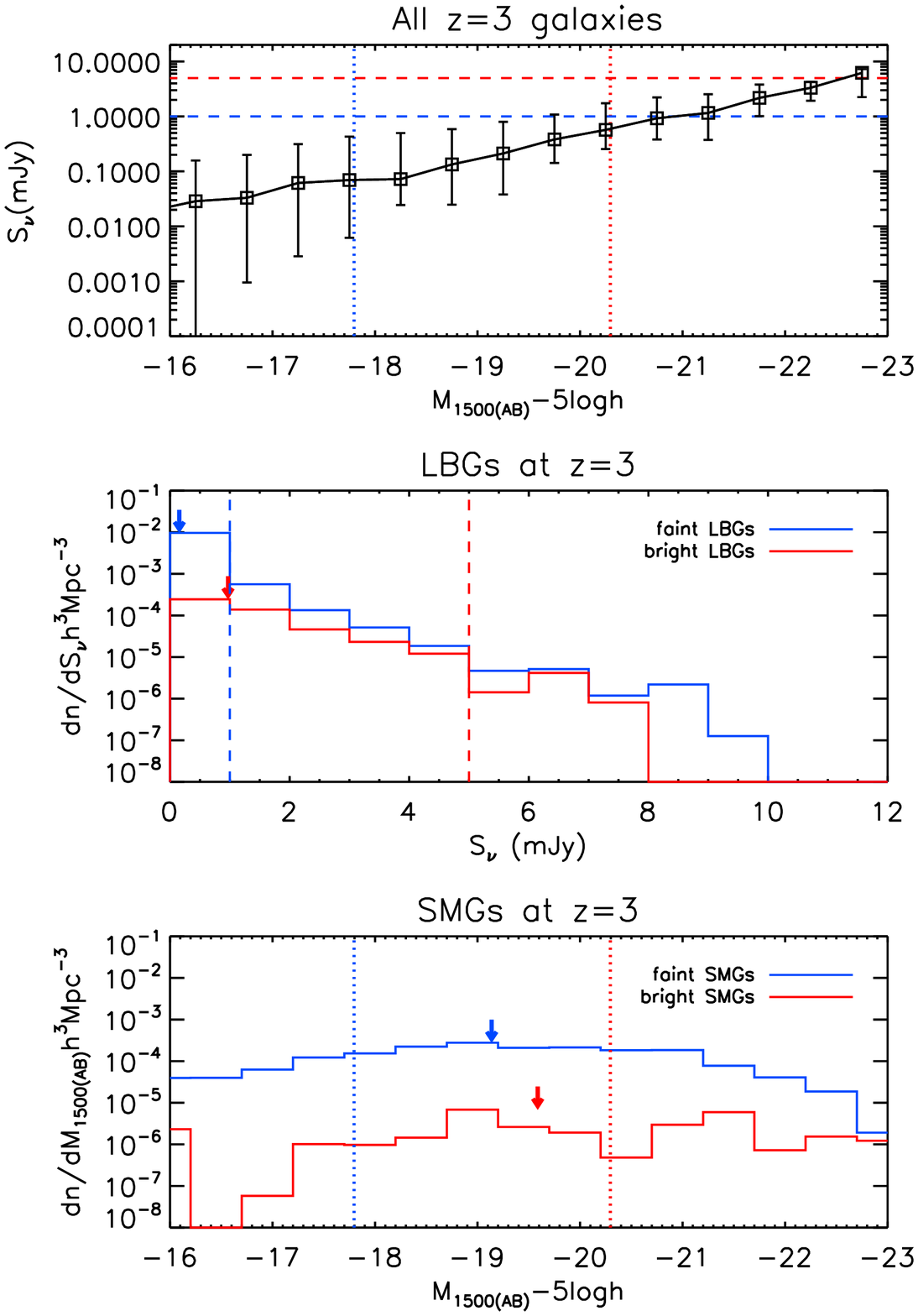}
\end{minipage}
\hspace{1cm}
\begin{minipage}{8cm}
\includegraphics[width=8cm, bb=55 35 550 760]{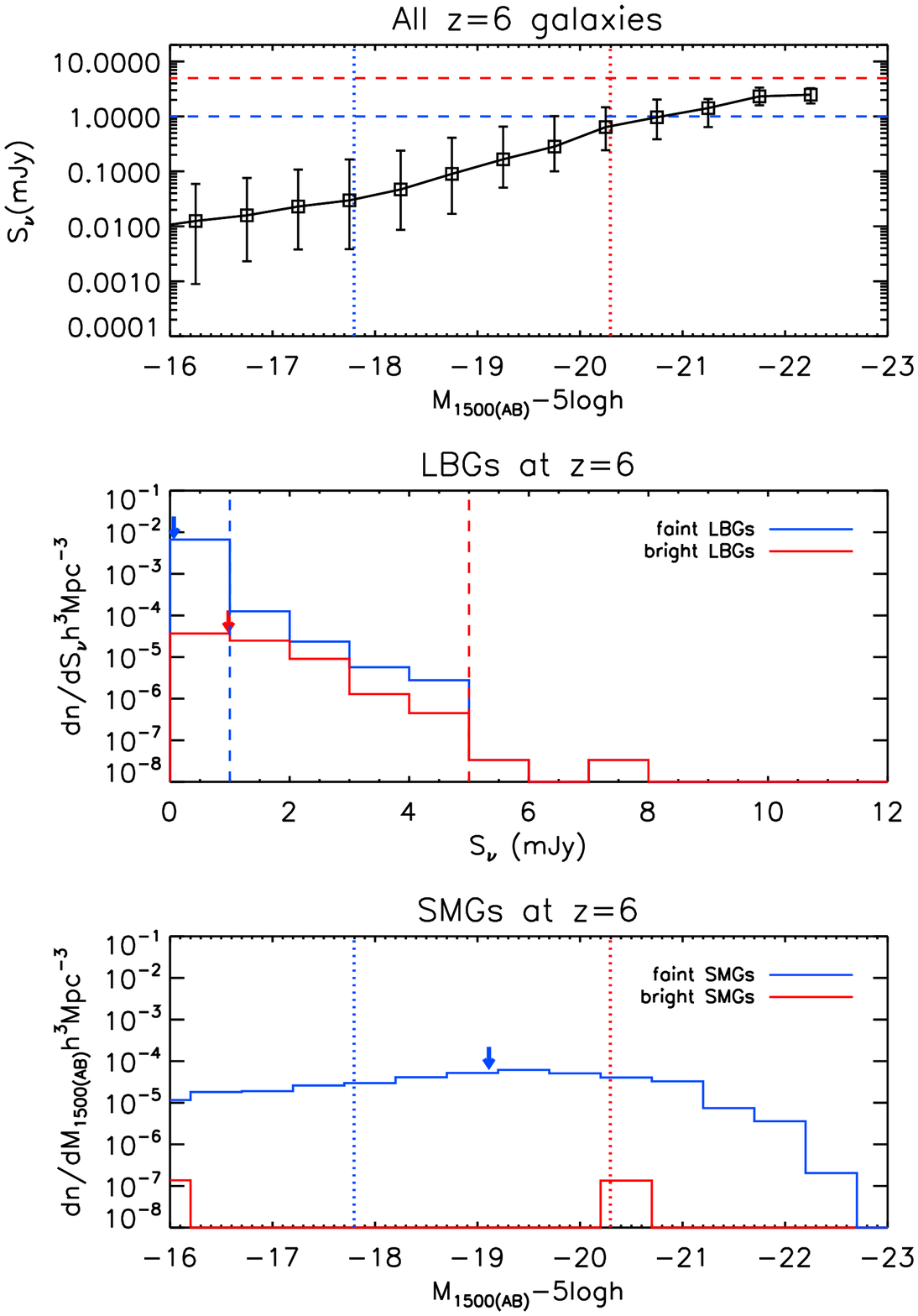}
\end{minipage}

\end{center}

\caption{ The relation between $850 ~\mu$m flux and rest-frame far-UV
  absolute magnitude for galaxies at $z=3$ (left panels) and $z=6$
  (right panels).  Top panels: the median $850 ~\mu$m flux as a
  function of the rest-frame far-UV absolute magnitude for galaxies at
  z=3. The bars indicate the 10 to 90 percentile range of the
  distribution in each magnitude bin.  The horizontal lines indicate 1
  and 5 mJy thresholds, which we used to define faint and bright SMGs
  in \citet{Gonzalez09b}, and the vertical lines indicate the faint
  ($0.1L^{*}_{UV}$) and bright ($L^{*}_{UV}$) LBG thresholds.  Middle
  panels: the number density of galaxies as a function of $850~\mu$m
  flux for faint LBGs (blue histogram) and bright LBGs (red
  histogram).  Bottom panels: the number density of galaxies as a
  function of UV absolute magnitude for faint SMGs ($S_{850} > 1 $mJy,
  blue histogram) and bright SMGs ($S_{850} > 5 $mJy, red
  histogram). In the middle and bottom panels, the arrows indicate the
  median value of the SMG flux and LBG luminosity, respectively, for
  each sample.  (We omit these arrows where there are too few galaxies
  in our model sample to reliably calculate the median.)}
\label{smglbg}

\end{figure*}

Samples of galaxies at high redshift have been assembled using the
Lyman-break technique and through their emission at sub-millimetre
wavelengths. LBGs are selected on the basis of their emission in the
rest-frame far-UV. SMGs are detected through their rest-frame
far-IR/sub-mm emission, which is produced by the reradiation by dust
of light absorbed predominantly in the UV. It is natural to ask what
is the overlap, if any, between these two populations at high
redshift. A first indication of the answer is suggested by the
different redshift ranges of the populations. LBGs selected using
ground-based surveys are restricted to have $z \gsim 3$. Bright SMGs
(with $850 \mu$m flux $S_{850} > 5~{\rm mJy}$), on the other hand,
have a median redshift of $z \sim 2$ in flux-limited samples. Hence,
to study the overlap between the two samples we will be looking at the
UV luminosities of galaxies in the high redshift tail of the SMG
redshift distribution.  \citet{Almeida10} performed a similar
comparison at $z=2$, and found that $\approx 1\%$ of galaxies bright
in the UV were predicted to also be bright SMGs.
 
The multiwavelength nature of our semi-analytical model and its
self-consistent treatment of dust make it ideally placed to address
the overlap between the SMG and LBG populations.  Here we examine the
predicted relation between $850 \mu {\rm m}$ flux and UV luminosity at
$z=3$ and $z=6$. Following the companion to this paper
\citep{Gonzalez09b}, in which we study the evolution of SMGs in the
model, we define a galaxy as a bright SMG when its $850 \mu$m flux
exceeds 5~mJy and as a faint SMG when its $850 \mu$m flux exceeds
1~mJy.

The top panels of Fig.~\ref{smglbg} show the relation between the
$850~\mu {\rm m}$ flux and the UV luminosity for galaxies at $z=3$ and
$z=6$, respectively. There is scatter in this relation, particularly
at fainter magnitudes. Nevertheless an increase in the UV luminosity
typically corresponds to an increase in the submillimetre flux.


The middle panels of Fig.~\ref{smglbg} show the number density of
galaxies as a function of their $850 \mu {\rm m}$ flux, with separate
curves for faint and bright LBGs at $z=3$ and $z=6$, respectively.  At
both redshifts, the median fluxes at $850 \mu$m are predicted to be
$1.0$mJy for bright LBGs, equal to the faint SMG flux limit adopted
here, but well below it ($\sim 0.1-0.2~$mJy) for faint LBGs. We find
that only 1.4\% of the bright LBGs ($L_{UV} > L^{*}_{UV}$) at $z=3$
are also bright SMGs ($S_{850} > 5 {\rm mJy}$). This figure decreases
to 0.1\% for $z=6$.

In the bottom panels of Fig.~\ref{smglbg}, we show the predicted
number density of galaxies as a function of UV absolute magnitude at
$z=3$ and $z=6$ respectively, distinguishing between faint and bright
SMGs.  At $z=3$, the median UV magnitude is quite similar for faint
and bright SMGs ($\sim 0.5 L^{*}$). At $z=6$ the median UV absolute
magnitude for the faint SMGs is predicted to be very similar to its
value at $z=3$. Our model predicts too few bright SMGs at $z=6$ for us
to reliably calculate the median UV absolute magnitude (as suggested
by the redshift distributions plotted in \citealt{Baugh05}).  At $z=3$,
nearly half (42\%) of the bright SMGs are predicted to also be bright
LBGs.

The model predicts that LBGs at $z=10$ should be extremely faint at
$850~\mu {\rm m}$, with faint LBGs having median fluxes of only
0.03~mJy. We predict a negligible number of bright or faint SMGs at
$z=10$.

\section[]{Summary}

We have investigated the nature, formation histories and present-day
descendants of Lyman-break galaxies (LBGs) in the $\Lambda$CDM
cosmology that are predicted by the \citet{Baugh05} version of the
{\tt GALFORM} semi-analytical galaxy formation model. The
\citeauthor{Baugh05} model incorporates a top-heavy IMF in starbursts
triggered by galaxy mergers, which was found to be necessary to
explain the number counts and redshift distribution of sub-mm
galaxies. This model has previously been shown to match the observed
far-UV luminosity function of LBGs over the whole range $z=3-10$
\citep{Lacey10b}. In this paper we investigate a range of other
predictions for Lyman-break galaxies from the same galaxy formation
model.


We began by investigating the formation histories of LBGs. We found
that the majority of LBGs are predicted to be starbursts triggered by
minor mergers of galaxies. Using the observed characteristic far-UV
luminosity at $z \approx 3$, $L^{*}_{UV}$, to define samples of model
galaxies as bright ($L_{UV}>L^{*}_{UV}$) or faint ($L_{UV}>0.1
L^{*}_{UV}$) LBGs, we found that $>99\%$ of bright LBGs at $z=3-6$ are
predicted to be starbursts triggered by minor or major mergers, while
this fraction is somewhat lower for faint LBGs (but still $>80\%$ for
$z=3-10$). The importance of starbursts in producing galaxies visible
as LBGs is because the far-UV light is dominated by young stars, and
in starbursts the star formation rate is increased over the quiescent
value. However, in our model there is also second important effect,
that the top-heavy IMF in bursts (as against a solar neighbourhood IMF
assumed for quiescent star formation) further enhances their far-UV
luminosities over quiescent galaxies.  We then examined the duration
of the LBG phase in model galaxies, and found that this is quite
short, around 20~Myr for bright LBGs, and somewhat longer, around
$20-60$~Myr, for faint LBGs. The short timescales are a result of the
LBGs mainly being starbursts.

We then computed the fraction of present-day galaxies that are
predicted to have LBG progenitors. We find that 95\% of Milky Way mass
galaxies had at least one faint LBG progenitor in the redshift range
$3<z<4$, and 70\% at $6<z<7$, but this falls to 7\% for $10<z<11$.
For bright LBG progenitors these fractions drop to 40\% for LBGs at
$3<z<4$ and only 3\% at $6<z<7$. Performing a similar analysis for
dark matter halos, we find that 50\% of present-day halos with masses
exceeding $4 \times 10^{12} \hMsol$ or $5 \times 10^{13} \hMsol$ had
progenitor halos hosting bright LBGs at $3<z<4$ or $6<z<7$
respectively.

We next examined the stellar and halo masses of LBGs, together with
their morphologies. The model predicts that bright LBGs at $z=3$ have
median stellar mass $\sim 1 \times 10^{9} \hMsol$, falling to $\sim 5
\times 10^{8} \hMsol$ at $z=6$, and are typically disk-dominated in
stellar mass, consistent with their origin in starbursts triggered by
minor galaxy mergers. As already noted by \citet{Lacey10b}, these
stellar masses are lower than the values typically inferred from
observational data by fitting stellar population models to broad-band
photometry, but the observational estimates rely on assuming a solar
neighbourhood IMF, while in our model the light from LBGs is dominated
by stars formed with a top-heavy IMF during a starburst. The faint
LBGs are predicted to have stellar masses roughly an order of
magnitude lower than bright LBGs at the same redshift, falling to only
$\sim 1 \times 10^{7} \hMsol$ at $z=10$, and to be more
bulge-dominated than the bright LBGs, especially at the higher
redshifts. The halos hosting bright LBGs at $z=3$ and $z=6$ are
predicted to have median masses $\sim  10^{11} \hMsol$, with roughly
10 times lower masses for faint LBGs at the same redshifts.

Following on from this, we then investigated the stellar and halo
masses and morphologies of the present-day descendants of LBGs. The
ability to relate different types of high-redshift galaxies to their
descendants is one of the unique capabilities of theoretical models of
galaxy formation such as ours. We find that the descendants of bright
LBGs at $z=3$ have median stellar masses $\approx 5 \times 10^{10}
\hMsol$ (i.e. comparable to the Milky Way), and consist of roughly
equal numbers of disk- and bulge-dominated systems. They typically
inhabit galaxy groups, with halo masses $\sim 10^{13} \hMsol$. Bright
LBGs at $z=6$ have descendants which are somewhat more massive
($\approx 1 \times 10^{11} \hMsol$) and inhabit somewhat more massive
halos, but are significantly more bulge-dominated ($\approx 75\%$) on
average. Faint LBGs are predicted to have present-day descendants
which are an order of magnitude lower in stellar and halo mass than
the bright LBGs at the same redshift, and contain a larger proportion
of disk-dominated systems.


We next looked at the relationship between halo mass, stellar mass,
star formation rate (SFR) and far-UV luminosity in the model, for all
galaxies and for LBGs. We find that LBGs have a tighter relationship
between halo mass and stellar mass than galaxies as a whole, due to
them them being mainly central galaxies. We also find a correlation
between SFR and stellar mass that is shifted to higher SFRs compared
to that for all galaxies. This results from LBGs being mainly
starbursts, which have higher SFRs than quiescent galaxies with the
same stellar mass. We also find that although most LBGs in the model
are starbursts, their stellar masses are not necessarily dominated by
stars formed in bursts. This results in part from the large recycled
fraction for stars forming with the top-heavy burst IMF. LBGs at $z=3$
(both faint and bright) are predicted to have formed most of their
stellar mass quiescently before the current burst. However, this
changes with redshift, so that LBGs at $z=6$ have typically formed
most of their stellar mass in bursts (although not necessarily the
current burst).

Finally we used the multi-wavelength capabilities of our model to
examine the overlap between LBGs and sub-mm galaxies (SMGs) detected
at $850 \mu$m, extending to $z=3$, 6 and 10 the comparison previously
carried out by \citet{Almeida10} at $z=2$. Analogously to the LBGs, we
define bright and faint SMGs as having $850 \mu$m fluxes exceeding
5~mJy and 1~mJy respectively. At $z \geq 3$ we are only probing the
tail of the redshift distribution of bright SMGs, which peaks at $z
\sim 2$. We find that only about 1\% of bright LBGs at $z=3$ are
predicted to be also bright SMGs, and this fraction falls to only
0.1\% for bright LBGs at $z=6$, with the median $850 \mu$m flux being
about 1~mJy for both redshifts. This explains the difficulty in
observationally detecting the sub-mm emission from LBGs
\citep{Chapman00}.

\section*{Acknowledgements} 
JEG acknowledges receipt of a fellowship funded by the European
Commission's Framework Programme 6, through the Marie Curie Early
Stage Training project MEST-CT-2005-021074.  This work was supported
in part by the Science and Technology Facilities Council rolling grant
to the ICC, and by the Austrian Science Foundation FWF under grant
I164. CSF acknowledges a Royal Society Wolfson Research Grant
Award. AJB acknowledges the support of the Gordon and Betty Moore
Foundation.

\bibliographystyle{mn2e_trunc8}
\bibliography{refs}

\bsp

\label{lastpage}

\end{document}